\documentclass[aps,prb,twocolumn,superscriptaddress]{revtex4}
\usepackage{graphicx}%
\usepackage{siunitx}
\usepackage{bm}
\usepackage[caption=false]{subfig}
\usepackage{mathtools}
\usepackage{multirow}
\usepackage{amssymb}
\usepackage{braket}
\usepackage{amsmath}
\usepackage{placeins}
\usepackage{cancel}
\usepackage{xcolor}
\usepackage{spreadtab}

\begin{document}

\title{Hole mobility of strained GaN from first principles}

\author{Samuel Ponc\'e}
\affiliation{%
Department of Materials, University of Oxford, Parks Road, Oxford, OX1 3PH, UK
}%
\author{Debdeep Jena}
\affiliation{%
School of Electrical and Computer Engineering, Cornell University, Ithaca, New York 14853, USA
}
\author{Feliciano Giustino}
\email{feliciano.giustino@materials.ox.ac.uk}
\affiliation{%
Department of Materials, University of Oxford, Parks Road, Oxford, OX1 3PH, UK
}%
\affiliation{%
Department of Material Science and Engineering, Cornell University, Ithaca, New York 14853, USA
}%

\date{\today}

\begin{abstract}
Nitride semiconductors are ubiquitous in optoelectronic devices such as LEDs and Blu-Ray optical disks. A major limitation for further adoption of GaN in power electronics is its low hole mobility. 
In order to address this challenge, here we investigate the phonon-limited mobility of wurtzite GaN using the \textit{ab initio} Boltzmann transport formalism,
including all electron-phonon scattering processes, spin-orbit coupling, and many-body quasiparticle band structures.
We demonstrate that the mobility is dominated by acoustic deformation-potential scattering, and
we predict that the hole mobility can significantly be increased by lifting the split-off hole states above the light and heavy holes. This can be achieved by reversing the sign of the crystal-field
splitting via strain or via coherent excitation the $A_1$ optical phonon through ultrafast infrared optical pulses.
\end{abstract}

\maketitle

\section{Introduction}

Wurtzite GaN plays a key role in solid state lighting
due its wide emission spectrum, high efficiency and scalable manufacturing~\cite{Zhou2017}.  
GaN has a high breakdown field, high thermal conductivity and high electron mobility, and is therefore an
excellent candidate for high power electronic devices~\cite{Ikeda2010,Ishida2016,Flack2016,Amano2018} and radio frequency electronics~\cite{Gassmann2007}.
GaN also exhibits high Seebeck coefficient and excellent temperature stability, which makes it a prime candidate for high temperature thermoelectric applications~\cite{Pantha2008,Sztein2009,Hurwitz2011}. 
It can also be used for thermal neutron and gamma radiation detection~\cite{Atsumi2014}.
More generally, group-III nitrides can be engineered to form continuous alloys with a tunable bandgap from 6.2~eV (AlN) through 3.4~eV (GaN) to 0.7~eV (InN)~\cite{Zhou2017},
finding applications in blue~\cite{Nakamura1997} and green~\cite{Lingrong2016} lasers, photodetectors~\cite{Sun2015}, 
and light-emitting diodes~\cite{Schubert2006}. 
Recently a GaN/NbN semiconductor/superconductor heterojunction was achieved through epitaxial growth, paving the way for superconducting qubits~\cite{Yan2018}.
Widespread adoption of GaN for applications such as complementary metal-oxide-semiconductor (CMOS) and high-power conversion devices is hindered
by the low hole mobility of GaN. 
In fact, the room temperature hole mobility of GaN does not exceed 40~cm$^2$/Vs~\cite{Kozodoy1998,Look1999,Rubin1994,Kozodoy2000,Cheong2000,Cheong2002,Horita2017}. 
In comparison, electron mobilities as high as 1265~cm$^2$/Vs have been reported in the bulk, and exceeding 2000~cm$^2$/Vs in 2D electron gases~\cite{Kyle2014}. 
It is therefore important to find practical ways to increase the hole mobility in this semiconductor. 

Transport properties in wurtzite GaN have been investigated theoretically decades ago by Ilegems and Montgomery~\cite{Ilegems1973}, taking into account conduction-band nonparabolicity,
deformation-potential, piezoelectric acoustic-phonon scattering, and polar-optical phonon scattering. 
More recently, analytical models based on experimental results
have been developed to accurately describe low-field carrier mobilities in a wide temperature and doping range~\cite{Mnatsakanov2003,Farahmand2001,Schwierz2005},
and the electron-phonon scattering rates of GaN were computed from first-principles using the EPW software~\cite{Jhalani2017}.

In a recent work~\cite{Ponce2019}, we clarified the atomic-scale mechanisms that are responsible for the low hole mobilities
in GaN using the state-of-the-art \textit{ab initio} Boltzmann transport formalism, and we discussed strategies to significantly increase the hole mobility in wurtzite GaN. 
We showed that the origin of the low \textit{hole} mobility lies in the scattering of carriers in the light-hole ($lh$) and heavy-hole
($hh$) bands  predominantly by long-wavelength longitudinal-acoustic phonons. 
Using this understanding, we predicted that the hole mobility could 
significantly be enhanced if the split-off hole band ($sh$) could be raised above the $lh$ and $hh$ bands. 
Such band inversion can be achieved by reversing the sign of the crystal-field splitting via uniaxial compressive or biaxial tensile strain.
In the present manuscript, we give details on the computational parameters used in Ref.~\onlinecite{Ponce2019}, convergences studies, phase diagram, and we discuss various strategies to perform momentum integration in the calculation of mobility. 

The manuscript is organized as follows.
%
%
In Sec.~\ref{phase_diag}, we compute the GaN phase diagram and show that the wurzite phase is thermodynamically stable for the range of pressures and temperatures investigated. 
We also discuss ground-state properties of unstrained GaN. 
We then discuss in Sec.~\ref{GWcorr} the electronic structure including many-body quasiparticle correction, the electron and hole effective masses as well as spin-orbit and crystal-field splitting.
In Sec.~\ref{carr_mob}, we briefly present the linearized Boltzmann transport formalism to compute the mobility of GaN, the associated convergence studies, the band velocity and the Hall factor.
We compare our theoretical results to experiment and analyze the origin of the low hole mobility in unstrained GaN. 
In Sec.~\ref{carr_mob_strain}, we analyze the elastic properties, phonon dispersion and electronic bandstructure of GaN under biaxial and uniaxial strain.
We also show how to increase the hole mobility via biaxial tensile or uniaxial compressive strain.
Finally, in Sec.~\ref{feasibility_gan}, we discuss how to realize high-mobility p-type GaN in experiments.
We draw our conclusions in Sec.~\ref{conclu}.

\section{Gallium nitride phase diagram and ground state properties}\label{phase_diag}

In this section we present the computed phase diagram of bulk GaN, highlighting the approximations and computational parameters used.
Such a diagram is important to make sure that no phase change occurs within the range of applied strains and temperature investigated in this study.  
We then focus on the stable wurtzite phase and discuss the optimized structure. 

GaN can form the following allotropes: (i) wurtzite P6$_3$mc, (ii) zincblende F$\bar{4}3$m and (iii) rock-salt Fm$\bar{3}$m.
The wurtzite phase is the naturally occurring phase, while the zincblende phase has been stabilized experimentally by epitaxial growth on cubic GaAs [001] surfaces~\cite{Brandt1995};
the rock-salt phase can be obtained under high pressure~\cite{Xia1993}.

\begin{figure}[b!]
  \centering
  \includegraphics[width=0.95\linewidth]{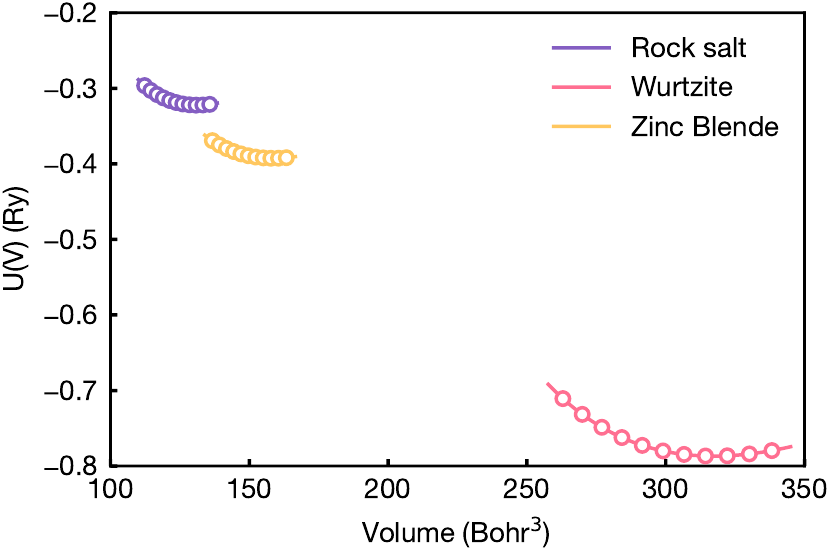}
  \caption{\label{figP1}
  Total energy versus volume at zero temperature for the rock salt, wurtzite and zinc blende phases of GaN. 
  }
\end{figure}

We use the fully relativistic norm-conserving Perdew-Zunger~\cite{Perdew1981} parameterization of the 
local density approximation (LDA) to density functional theory, and the Perdew-Burke-Ernzerhof~\cite{Perdew1996} generalized gradient approximation (PBE). 
The pseudopotentials are generated using the ONCVPSP 
code~\cite{Hamann2013} and optimized via the PseudoDojo initiative~\cite{Setten2018}. 
The semicore $3s$, $3p$, and $3d$ electrons of Ga are explicitly described, as GW quasiparticle corrections are sensitive
to semicore states. 
The electron wavefunctions are expanded in a planewave basis set with kinetic energy cutoff of 120~Ry, and the Brillouin zone is
sampled using an homogeneous $\Gamma$-centered 6$\times$6$\times$6 mesh. 

\begin{figure}[t!]
  \centering
  \includegraphics[width=0.99\linewidth]{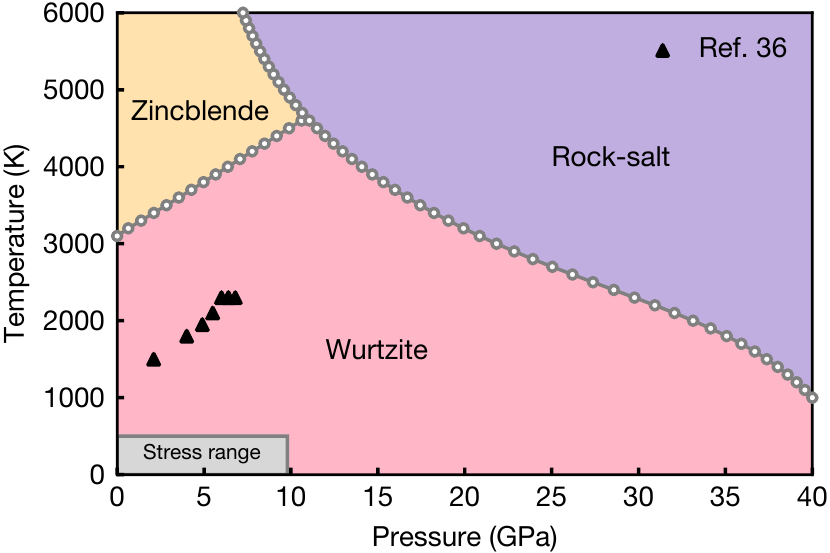}
  \caption{\label{figP2}
  Phase diagram of GaN. The phonon frequencies are computed using the PBE functional, without spin-orbit coupling, at 11 different volumes. 
  The experimental values (black triangles) are from Ref.~\onlinecite{Utsumi2003}. 
  The light gray rectangle represents the maximum strain (9.8~GPa at 2\% strain) and temperature (500~K) investigated in this paper.
  }
\end{figure}

\begin{figure*}[t!]
  \centering
  \includegraphics[width=0.99\linewidth]{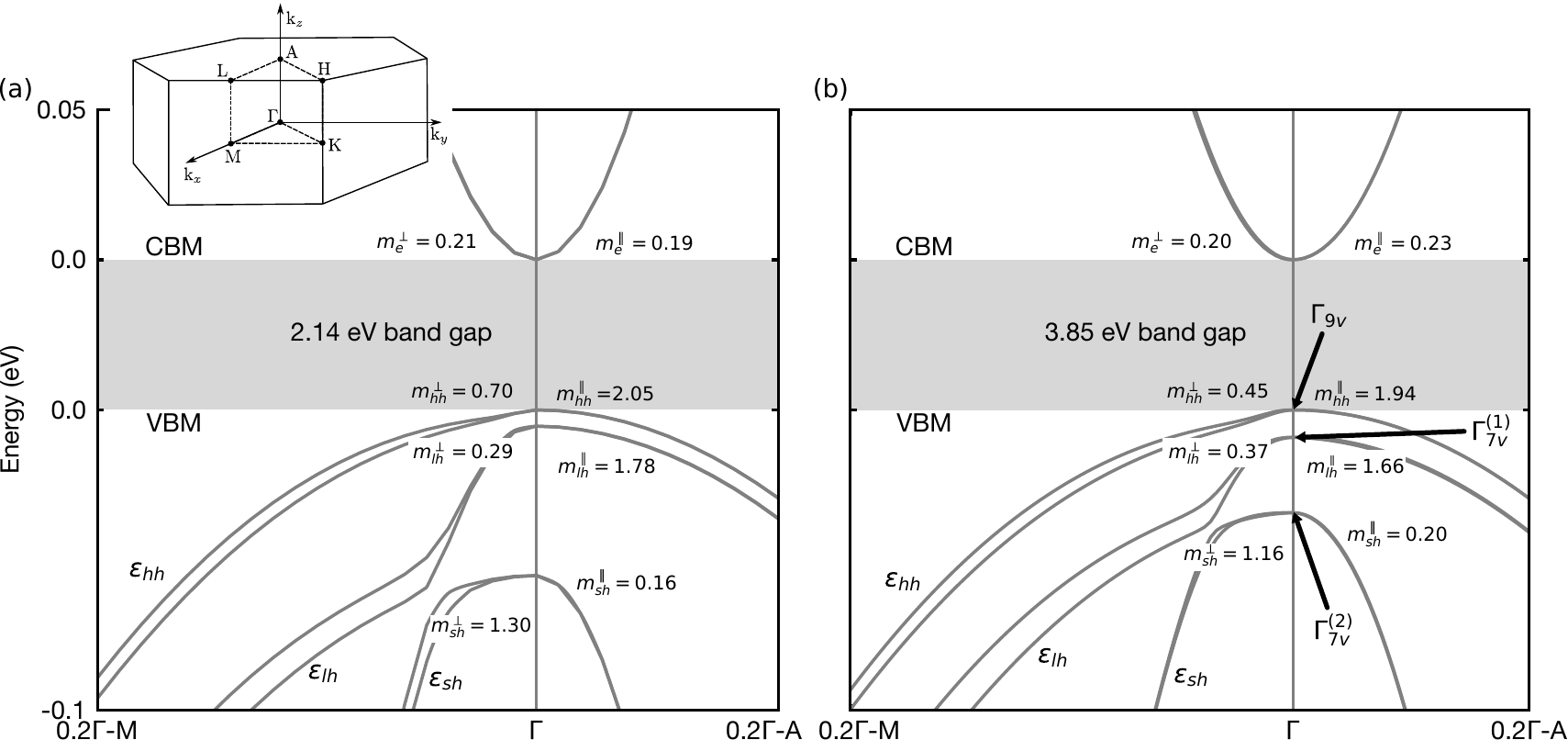}
  \caption{\label{figP3}
  Electronic bandstructure of wurtzite GaN using (a) the LDA functional in the optimized ground-state LDA structure, and 
  (b) quasiparticle G$_0$W$_0$+$\Delta_{\mathbf{k}}$ calculation.
  We indicate the effective masses at the zone center, obtained from the second derivatives of the
  band energy with respect to the wavevector along the $\Gamma$M and $\Gamma$A directions, 
  respectively. The bandgap is off scale for clarity. 
  We indicate the naming convention for the three topmost eigenstates at $\Gamma$. 
  The energy levels have been aligned to the band edges.
  A schematic of the Brillouin zone of wurtzite GaN is given in the upper left corner.
  }
\end{figure*}

Within the quasi-harmonic approximation~\cite{Baroni2010}, the Helmholtz free energy of a crystal is given by~\cite{Palumbo2017}:
\begin{equation}
F(T,V) = U(V) + F^{\rm{vib}}(T,V) + F^{\rm{el}}(T,V),
\end{equation}
where $U$ is the static (clamped-ion) energy at 0~K, $F^{\rm{vib}}$ is the contribution due to lattice vibrations 
and $F^{\rm{el}}$ is the energy due to electronic thermal excitations. 
We rely on the adiabatic approximation to treat each term independently. 
The vibrational Helmholtz free energy per cell is given in the harmonic approximation by~\cite{Palumbo2017}: 
\begin{multline}
F^{\textrm{vib}}(T,V) = \frac{1}{2N} \sum_{\mathbf{q},\nu} \hbar \omega_{\mathbf{q},\nu}(V) \\
+ \frac{k_B T}{N}  \sum_{\mathbf{q},\nu} \ln \bigg[ 1 - \exp\Big( \frac{-\hbar \omega_{\mathbf{q},\nu}(V)}{k_B T} \Big) \bigg],
\end{multline}
where $N$ is the number of $\bf q$-points, the first term is the contribution to the zero-point energy and the second term is the 
phonon contribution at finite temperature. $F^{\rm{el}}$ can be neglected as the band
gap is much larger than thermal energies.

The energy minimum of $U(V) + F^{\rm{vib}}(T,V)$ at a given temperature corresponds to zero pressure and gives the variation of volume
with temperature due to thermal expansion. 
To perform these calculations we use the Quantum Espresso~\cite{Giannozzi2017} and thermo\_pw codes~\cite{Corso2016}.
The phonon frequencies were computed using the PBE pseudopotentials, without spin-orbit coupling (SOC), at 11 different volumes. 
The resulting energies were fitted using the Murnaghan equation of state~\cite{Murnaghan1944}. 
We used a 6$\times$6$\times$6 $\mathbf{q}$-point grid for the phonons.  
In order to obtain accurate dielectric permittivity tensors and Born effective charges
we employed a much denser, shifted Monkhorst-Pack grid with 16$\times$16$\times$16 $\mathbf{k}$-points.
To compute phonon dispersion relations, we applied the crystal acoustic sum rule~\cite{Mounet2005,Mingo2008}.
The calculated dependence of the static energy $U(V)$ on the volume is shown in Fig.~\ref{figP1}.

The complete phase diagram can be obtained by comparing the Gibbs free energy of the various allotropes at each temperature and pressure. 
The Gibbs free energy $G(T,P)$ can be obtained from the Helmholtz free energy as:
\begin{equation}
G(T,P) = F(T,V) + PV,
\end{equation}
where the pressure is obtained by computing the first-order derivative of the Helmholtz free energy, with respect to volume at fixed temperature:
\begin{equation}
P = -\frac{\partial F}{\partial V}\bigg|_T.
\end{equation} 
The Gibbs free energy was computed using the python toolkit PhaseGO~\cite{Liu2015}.
The resulting pressure-temperature phase diagram is given in Fig.~\ref{figP2}, where the light gray rectangle represents the maximum stress (9.8~GPa at 2\% strain) and temperature (500~K) investigated in this paper.
We see that the wurtzite structure is always the thermodynamically stable phase in the strain and temperature considered in this study.
We note that our calculated \textit{ab initio} phase diagram overestimates the coordinates of the triple point (10~GPa, 4700~K) with respect to experiment (6.2~GPa, 2300~K)~\cite{Utsumi2003}, however this error does not affect the region of interest for our present study.

Our optimized lattice parameters of wurtzite GaN are $a=5.961 (6.081)$~bohr and $c=9.716 (9.9049)$~bohr, and the internal parameter is $u=0.376 (0.377)$ in LDA (PBE). 
As expected, the experimentally measured parameters, $a=6.026$~bohr and $c=9.800$~bohr~\cite{Qian1996}, fall in between the LDA and PBE data.
All subsequent calculations of electron band structures, phonon dispersion relations, and electron-phonon interactions are performed using these optimized lattice parameters.

\section{GW quasiparticle corrections}\label{GWcorr}

\begin{table}[b]
  \begin{tabular}{l c c c c c c c c}
  \toprule 
                                     & \multicolumn{8}{c}{Energy gaps} \\
                                     & \multicolumn{2}{c}{E$_g$ } &  & \multicolumn{2}{c}{$\Delta_{\rm so}$ } & & \multicolumn{2}{c}{$\Delta_{\rm cf}$ } \\
\rule{0pt}{1em}Present work          & \multicolumn{2}{c}{(eV)} & & \multicolumn{2}{c}{(meV)} & & \multicolumn{2}{c}{(meV)} \\
\hline
PBE                                  & \multicolumn{2}{c}{1.74}  &  & \multicolumn{2}{c}{9}    & & \multicolumn{2}{c}{39} \\
PBE\!+\!G$_0$\!W$_0$                       & \multicolumn{2}{c}{2.60}  &  & \multicolumn{2}{c}{11}   & & \multicolumn{2}{c}{26} \\
PBE\!+\!G$_0$\!W$_0$\!+\!$\Delta_{\mathbf{k}}$ & \multicolumn{2}{c}{2.94}  &  & \multicolumn{2}{c}{13}   & & \multicolumn{2}{c}{22} \\
LDA                                  & \multicolumn{2}{c}{2.14}  &  & \multicolumn{2}{c}{8}    & & \multicolumn{2}{c}{53} \\
LDA\!+\!G$_0$\!W$_0$                       & \multicolumn{2}{c}{3.41}  &  & \multicolumn{2}{c}{12}   & & \multicolumn{2}{c}{35} \\
LDA\!+\!G$_0$\!W$_0$\!+\!$\Delta_{\mathbf{k}}$ & \multicolumn{2}{c}{\textbf{3.85}}  &  & \multicolumn{2}{c}{\textbf{14}}   & & \multicolumn{2}{c}{\textbf{30}} \\
\rule{0pt}{1em}Previous work                       & & & & & & & & \\
\hline
LDA~\cite{Wei1996}                   & \multicolumn{2}{c}{ -  }  &  & \multicolumn{2}{c}{13}   & & \multicolumn{2}{c}{42} \\
LDA~\cite{Chen1996}                  & \multicolumn{2}{c}{ -  }  &  & \multicolumn{2}{c}{12}   & & \multicolumn{2}{c}{37} \\
LDA~\cite{Kim1997}                   & \multicolumn{2}{c}{ -  }  &  & \multicolumn{2}{c}{16}   & & \multicolumn{2}{c}{36} \\
LDA~\cite{Rinke2008}                 & \multicolumn{2}{c}{1.78}  &  & \multicolumn{2}{c}{-}    & & \multicolumn{2}{c}{49} \\
OEPx\!+\!G$_0$\!W$_0$\cite{Rinke2008}& \multicolumn{2}{c}{3.24}  &  & \multicolumn{2}{c}{-}    & & \multicolumn{2}{c}{34} \\
QSGW~\cite{Svane2010}                & \multicolumn{2}{c}{3.81}  &  & \multicolumn{2}{c}{-}    & & \multicolumn{2}{c}{-} \\
0.8$\Sigma$-QSGW~\cite{Svane2010}    & \multicolumn{2}{c}{3.42}  &  & \multicolumn{2}{c}{-}    & & \multicolumn{2}{c}{-} \\
0.8$\Sigma$-QSGW~\cite{Punya2012}    & \multicolumn{2}{c}{3.60}  &  & \multicolumn{2}{c}{5$^\ddagger$}    & & \multicolumn{2}{c}{18$^\ddagger$} \\
0.8$\Sigma$-QSGW~\cite{Punya2012}    & \multicolumn{2}{c}{3.60}  &  & \multicolumn{2}{c}{14}   & & \multicolumn{2}{c}{12} \\
Experiment~\cite{Monemar1974}        & \multicolumn{2}{c}{3.47}  &  & \multicolumn{2}{c}{-}    & & \multicolumn{2}{c}{-} \\
Experiment~\cite{Dingle1971}         & \multicolumn{2}{c}{3.47}  &  & \multicolumn{2}{c}{11}   & & \multicolumn{2}{c}{22} \\
Experiment~\cite{Gil1995}            & \multicolumn{2}{c}{-}     &  & \multicolumn{2}{c}{18}   & & \multicolumn{2}{c}{10} \\
Experiment~\cite{Chuang1996}         & \multicolumn{2}{c}{-}     &  & \multicolumn{2}{c}{12}   & & \multicolumn{2}{c}{16} \\
Experiment~\cite{Reynolds1996}       & \multicolumn{2}{c}{3.51}  &  & \multicolumn{2}{c}{17}   & & \multicolumn{2}{c}{25} \\
Experiment~\cite{Rodina2001}         & \multicolumn{2}{c}{-}     &  & \multicolumn{2}{c}{19}   & & \multicolumn{2}{c}{10} \\
\\
& \multicolumn{8}{c}{Effective masses} \\  
                          & m$_{hh}^{\parallel}$   & m$_{lh}^{\parallel}$ & m$_{sh}^{\parallel}$ & m$_{hh}^{\perp}$ & m$_{lh}^{\perp}$ & m$_{sh}^{\perp}$ & m$_{e}^{\parallel}$ & m$_{e}^{\perp}$\\  
\rule{0pt}{1em}Present work          & & & & & & & & \\
\hline
LDA                                  & 2.05 &                 1.78 &                 0.16 &             0.70 &             0.29 &             1.30 &               0.19  &      0.21    \\
LDA\!+\!G$_0$\!W$_0$                       & 1.98 &                 1.68 &                 0.18 &             0.58 &             0.33 &             1.29 &               0.22  &      0.19    \\
LDA\!+\!G$_0$\!W$_0$\!+\!$\Delta_{\mathbf{k}}$ & \textbf{1.94} &\textbf{1.66} &       \textbf{0.20} &     \textbf{0.45}&     \textbf{0.37}&     \textbf{1.16}&       \textbf{0.23} &\textbf{0.20} \\
\rule{0pt}{1em}Previous work          & & & & & & & & \\
\hline
Theory~\cite{Punya2012}              & 1.85 &                 0.55 &                 0.20 &             0.69 &             0.50 &             0.80 &               0.20  &      0.22    \\
Theory*~\cite{Rinke2008}             & 1.88 &                 0.92 &                 0.19 &             0.33 &             0.36 &             1.27 &               0.19  &      0.21    \\ 
Theory$^{\dagger}$~\cite{Rinke2008}  & 1.88 &                 0.37 &                 0.26 &             0.33 &             0.49 &             0.65 &               0.19  &      0.21    \\
Theory~\cite{Kim1997}                & 2.00 &                 1.19 &                 0.17 &             0.34 &             0.35 &             1.27 &               0.19  &      0.23    \\   
Theory~\cite{Chen1996}               & 2.03 &                 1.25 &                 0.15 &             0.33 &             0.34 &             1.22 &               0.17  &      0.19    \\      
Experiment~\cite{Rodina2001}         & 1.76 &                 0.42 &                 0.30 &             0.35 &             0.51 &             0.68 &                 -   &        -     \\
Experiment~\cite{Feneberg2013}       &   -  &                   -  &                  -   &               -  &               -  &               -  &               0.22  &      0.24   \\
  \botrule 
  \end{tabular}
  \caption{\label{table1}
  Comparison between our calculated bandgap E$_g$, spin-orbit splitting $\Delta_{\rm so}$, crystal-field splitting $\Delta_{\rm cf}$, and effective masses of wurzite GaN with 
  earlier theory and experiment. QSGW stands for quasiparticle self-consistent approach~\cite{Schilfgaarde2006}, 0.8$\Sigma$-QSGW is an empirical hybrid method with 20\% LDA self-energy~\cite{Chantis2006} and OEPx stands for exact-exchange optimized effective potential~\cite{Rinke2008}.  
  $^{\ddagger}$ Calculation done using the quasicubic approximation.
  * Calculation using $\Delta_\text{so}=16$~meV and $\Delta_{\text{cf}}=25$~meV from Ref.~\onlinecite{Carrier2005}. 
  $^{\dagger}$ Calculation using  $\Delta_\text{so}=19$~meV and $\Delta_{\text{cf}}=10$~meV from Ref.~\onlinecite{Rodina2001}. 
  The bold values are recommended and are used throughout this manuscript.     
  }
\end{table}

The electronic bandstructure computed within the LDA is presented in Fig.~\ref{figP3}(a) for states close to the band edge along the high-symmetry directions $\Gamma$-A (which is the direction parallel to the $c$-axis, denoted with a $\parallel$ symbol) and $\Gamma$-M (which is the perpendicular direction, denoted with a $\perp$ symbol). 
Due to the wurtzite symmetry, the in-plane $\Gamma$-K direction is equivalent to the $\Gamma$-M direction, and therefore is not shown.

The calculated direct bandgap is 2.14~eV, strongly underestimating the measured value of 3.5~eV~\cite{Monemar1974,Vurgaftman2003}.  
To overcome this shortcoming, we calculated the GW quasiparticle band structures of wurtzite GaN within the many-body G$_0$W$_0$ approximation including 
SOC, as implemented into the Yambo code~\cite{Marini2009}. 
We used a planewaves kinetic energy cutoffs of 120~Ry for the exchange self-energy and 29~Ry for the polarizability.
In addition, we included 1500 bands, a plasma energy of 16.5~eV for the plasmon pole~\cite{Godby1989}, and a
6$\times$6$\times$6 $\Gamma$-centered Brillouin-zone grid. 
We employed the band extrapolation scheme of Ref.~\onlinecite{Bruneval2008} to speed-up convergence with the number of empty states. 
We obtained a corrected bandgap of 3.41~eV, much closer to the experimental one.

As discussed in Ref.~\onlinecite{Ponce2019a}, the accuracy of the effective masses is improved by using a self-consistent, ${\bf k}$-dependent scissor shift. 
This approximation, which we call G$_0$W$_0$-$\Delta_{\mathbf{k}}$, yields a wider bandgap of 3.85~eV. 
This value is in agreement with the gaps 3.24~eV and 3.81~eV obtained in previous calculations~\cite{Rinke2008,Svane2010}. 
The calculated LDA+G$_0$W$_0$+$\Delta_{\mathbf{k}}$ band structure for unstrained GaN is reported in Fig.~\ref{figP3}(b).
Throughout this paper, GW-corrected band structures were obtained via Wannier interpolation~\cite{Marzari2012}, using 20 Wannier functions
for the ground-state structure and the structures with 1\% strain, and 28 Wannier functions for the 
structures with 2\% strain (to be discussed below).
In contrast, we note that the PBE functional yields much too small a bandgap, even after G$_0$W$_0$+$\Delta_{\mathbf{k}}$ corrections (2.94~eV).
The various bandgaps and their comparison to previous calculations and experiment are summarized in Table~\ref{table1}.
We note that the theoretical values reported in Table~\ref{table1} do not account for the zero-point renormalization, which has been calculated to be -150~meV for zinc-blende GaN~\cite{Nery2016}.

As shown in Fig.~\ref{figP3}, the conduction band bottom of GaN is singly degenerated, while 
at the valence band top we have a $lh$ and a $hh$, which are split into 
doublets by SOC as we move from the $\Gamma$ to the M point of the Brillouin zone. 
We also have a split-off hole resulting from crystal-field splitting.
In order to determine the spin-orbit splitting $\Delta_{\rm so}$ and the crystal-field splitting
$\Delta_{\rm cf}$, we employ the quasi-cubic model of Refs.~\onlinecite{Wei1996,Hopfdeld1960}
for the triplet of states $\Gamma_{9v}$, $\Gamma_{7v}^{(1)}$, and $\Gamma_{7v}^{(2)}$ at the valence 
band top, with energies $\varepsilon_{hh}$, $\varepsilon_{lh}$, and $\varepsilon_{sh}$, respectively:
\begin{align}
  \varepsilon_{hh}                  &= \phantom{-}\frac{1}{2}(\Delta_{\rm so} + \Delta_{\rm cf}),  \\
  \varepsilon_{lh},\varepsilon_{sh} &= \pm\frac{1}{2}\sqrt{(\Delta_{\rm so} + \Delta_{\rm cf})^2 - \frac{8}{3}\Delta_{\rm so} \Delta_{\rm cf}}.  
\end{align}
Having checked that $\Delta_{\rm so}\ll \Delta_{\rm cf}$, we can simplify these expressions as:
\begin{align}\label{delat_so}
  \Delta_{\rm so} &= \frac{3}{2}(\varepsilon_{hh}-\varepsilon_{lh}), \\
  \Delta_{\rm cf} &= \varepsilon_{lh}-\varepsilon_{sh} +\frac{\varepsilon_{hh}-\varepsilon_{lh}}{2}.\label{delat_cf}
\end{align}
Using the last two equations, we determine the spin-orbit splitting and the
crystal-field splitting from the calculated band structure energies $\varepsilon_{hh}$, $\varepsilon_{lh}$, and $\varepsilon_{sh}$ and
systematically report them in Table~\ref{table1}.
As seen in Fig.~\ref{figP3}, the effect of G$_0$W$_0$ and the self-consistent scissor is to increase the spin-orbit splitting and decrease 
the crystal-field splitting. 
The same effect is observed using either the LDA or PBE exchange-correlation functional. 
In our LDA+G$_0$W$_0$+$\Delta_{\mathbf{k}}$ calculations for unstrained GaN we find $\Delta_\text{so}=14$~meV and 
$\Delta_{\text{cf}}=30$~meV, in the range of experimental values 
$\Delta_\text{so}=$11-19~meV and $\Delta_\text{cf}=$10-25~meV~\cite{Dingle1971,Chuang1996,Chen1996,Shikanai1997,Reynolds1996,Rodina2001}.
We note that LDA tends to slightly overestimate the crystal-field splitting with respect to experiment, in line with previous theoretical findings~\cite{Wei1996,Chen1996,Kim1997,Rinke2008,Svane2010}.
In contrast, PBE yields slightly smaller values for the crystal-field splitting, but since the bandgap is strongly underestimated, we proceed with LDA for the remainder of the paper. 

Using the parabolic band approximation, our calculated effective masses with quasiparticle and scissor-shift corrections are $m_{e}^{\perp/\parallel} = 0.20/0.23\,m_{\rm e}$, $m_{hh}^{\perp/\parallel} = 0.45/1.94\,m_{\rm e}$, $m_{lh}^{\perp/\parallel} = 0.37/1.66\,m_{\rm e}$, and  $m_{sh}^{\perp/\parallel} = 1.16/0.2\,m_{\rm e}$, respectively.
These values are in reasonable agreement with experimental data ranging 
from $0.30\,m_{\rm e}$ to $2.03\,m_{\rm e}$~\cite{Pankove1975,Xu1993, Fan1996,Yeo1998,Rodina2001} for holes, 
and in good agreement with $0.2\,m_{\rm e}$~\cite{Drechsler1995} for the electrons. 
A detailed comparison with previously computed effective masses and experimental masses is given in Table~\ref{table1}.
The largest discrepancy with respect to experiment are the overestimated m$_{lh}^{\parallel}$ and m$_{sh}^{\perp}$  effective masses, resulting 
from an overestimation of the crystal-field splitting in the LDA. 
This effect was already reported in Ref.~\onlinecite{Rinke2008}.
Indeed, as seen in Table~\ref{table1}, their m$_{lh}^{\parallel}$ and m$_{sh}^{\perp}$ effective masses
decrease from 0.92 to 0.37 and from 1.27 to 0.65 when using the calculated $\Delta_{\text{cf}}=25$~meV from Ref.~\onlinecite{Carrier2005}
or the experimental value of $\Delta_\text{cf}=10$~meV from Ref.~\onlinecite{Rodina2001}. 
Overall we find that increasing the level of theory (from LDA to LDA+G$_0$W$_0$ to LDA+G$_0$W$_0$+$\Delta_{\mathbf{k}}$) systematically improves all the effective masses with respect to the experimental values.
Closer agreement with experiment could be achieved by including the small effect of polaronic mass enhancement to the \textit{ab-initio} calculations~\cite{Lambrecht2017}.
We also note that the electron effective mass has been confirmed by quantum magnetotransport measurements~\cite{Jena2003,Knap2004}, but a corresponding high accuracy measurement has not been achieved yet for holes in GaN.

\section{Carrier mobility in unstrained GaN}\label{carr_mob}
\subsection{Linearized Boltzmann transport equation}

The carrier drift mobility $\mu$ describes the change of steady-state carrier current $J_\alpha = e(n_{\rm e} \mu_{{\rm e},\alpha\beta} + n_{\text{h}}\mu_{{\rm h},\alpha\beta})E_\beta$ due to an applied external electric field $\mathbf{E}$, where Greek indices
denote Cartesian coordinates, $n_{\rm e}$ and $n_{\rm h}$ the electron and hole density, respectively. 
The mobility can be computed using the linearized Boltzmann transport equation 
(BTE)~\cite{Ziman1960,Kaasbjerg2012a,Li2015,Fiorentini2016,Zhou2016,Gunst2016,Ponce2018,Ma2018,Macheda2018}, which for electrons reads:
  \begin{equation}\label{eq.1}
  \mu_{{\rm e},\alpha\beta} = \frac{-1}{n_{\rm e} \Omega} \sum_{n\in {\rm CB}}
  \int \frac{d\mathbf{k}}{\Omega_{\rm BZ}}  v_{n\mathbf{k},\alpha} \partial_{E_\beta} f_{n\mathbf{k}}.
  \end{equation}
Here $v_{n\mathbf{k},\alpha} = \hbar^{-1}\partial \varepsilon_{n\mathbf{k}}/\partial k_\alpha$ is 
the group velocity of the band state of energy $\varepsilon_{n\mathbf{k}}$, band index $n$, and wavevector $\bf k$. 
${\rm CB}$ stands for conduction bands, $\partial_{E_\beta} f_{n\mathbf{k}}$ is the perturbation to the 
Fermi-Dirac distribution induced by the applied electric field $\mathbf{E}$; $\Omega$ and $\Omega_{\rm BZ}$ 
are the volumes of the crystalline unit cell and first Brillouin zone, respectively. 
We note that the additional term in the velocity arising from the Berry curvature contribution
vanishes in bulk GaN due to time-reversal symmetry and does not contribute to the BTE mobility~\cite{Xiao2010}.
The perturbation to the equilibrium carrier distribution is obtained by solving 
the following self-consistent equation:
  \begin{multline}\label{eq.2}
  \partial_{E_\beta} f_{n\mathbf{k}} = e \frac{\partial f^0_{n\mathbf{k}}}{\partial 
  \varepsilon_{n\mathbf{k}}} 
  v_{n\mathbf{k},\beta} \tau_{n\mathbf{k}}  \!+ \frac{2\pi\tau_{n\mathbf{k}}}{\hbar} \sum_{m\nu} \!\int\!\! \frac{d\mathbf{q}}{\Omega_{\text{BZ}}} | g_{mn\nu}(\mathbf{k,q})|^2  \\
   \times \big[ ( n_{\mathbf{q}\nu} + 1 - f_{n\mathbf{k}}^0 ) \delta(\Delta \varepsilon^{nm}_{\mathbf{k},\mathbf{q}} + \hbar \omega_{\mathbf{q}\nu} ) \\
   + ( n_{\mathbf{q}\nu} +  f_{n\mathbf{k}}^0 )\delta(\Delta \varepsilon^{nm}_{\mathbf{k},\mathbf{q}} - \hbar \omega_{\mathbf{q}\nu} )   \big]
  \partial_{E_\beta} f_{m\mathbf{k+q}},
  \end{multline}
where $\Delta \varepsilon^{nm}_{\mathbf{k},\mathbf{q}} =
\varepsilon_{n\mathbf{k}} - \varepsilon_{m\mathbf{k+q}}$, $f^0_{n\mathbf{k}}$ is the equilibrium 
distribution function, and $n_{\mathbf{q}\nu}$ is the Bose-Einstein occupation. 
The matrix elements $g_{mn\nu}(\mathbf{k}, \mathbf{q})$ are the probability 
amplitudes for scattering from an initial electronic state $n{\bf k}$ to a final state $m{\bf k+q}$
via a phonon of branch index $\nu$, crystal momentum $\bf q$, and frequency $\omega_{\mathbf{q}\nu}$:
\begin{equation}\label{gmat}
	g_{mn,\nu}(\mathbf{k,q}) = \Big[\frac{\hbar}{2 M_{\kappa} \omega_{\mathbf{q}\nu}}\Big]^{1/2} \langle \psi_{m\mathbf{k+q}}  | \partial_{\mathbf{q}\nu}V | \psi_{n\mathbf{k}} \rangle,
\end{equation}
where $M_\kappa$ is the mass of the atom $\kappa$ and $\partial_{\mathbf{q}\nu}V$ is the derivative of the self-consistent potential associated with a phonon of wavevector \textbf{q}. 
$\psi_{n\mathbf{k}}$ is the electronic wavefunction for band $n$ and wavevector $\mathbf{k}$. 

The quantity $\tau_{n\mathbf{k}}$ in Eq.~\eqref{eq.2} is the relaxation time, and is given by~\cite{Grimvall1981,Giustino2017}:
\begin{multline}\label{eq.3}
  \frac{1}{\tau_{n\mathbf{k}}} = \frac{2\pi}{\hbar} \sum_{m\nu\sigma}\int \frac{d\mathbf{q}}{\Omega_{\rm BZ}}  
  |g_{mn\nu}(\mathbf{k},\mathbf{q})|^2  \\ 
   \times \big[ ( n_{\mathbf{q}\nu} + 1 - f_{m\mathbf{k+q}}^0 ) \delta(\Delta \varepsilon^{nm}_{\mathbf{k},\mathbf{q}} - \hbar \omega_{\mathbf{q}\nu} ) \\
   + ( n_{\mathbf{q}\nu} +  f_{m\mathbf{k+q}}^0 )\delta(\Delta \varepsilon^{nm}_{\mathbf{k},\mathbf{q}} + \hbar \omega_{\mathbf{q}\nu} )   \big]  
\end{multline}
In our calculations we first compute Eq.~(\ref{eq.3}), then solve Eq.~(\ref{eq.2}) iteratively to
obtain $\partial_{E_\beta} f_{n\mathbf{k}}$, and we use the result inside Eq.~(\ref{eq.1}).

A common approximation for calculating mobilities is to neglect the second term on the right-hand side
of Eq.~\eqref{eq.2}. In this case the relaxation time is explicitly given by Eq.~\eqref{eq.3}, and the equations are solved non-self-consistently. 
In Ref.~\onlinecite{Ponce2018} we called 
this simplification the self-energy relaxation time approximation (SERTA), and the corresponding mobility is given explicitly by:
\begin{equation}\label{SERTAeq}
  \mu_{{\rm e},\alpha\beta}^{\text{SERTA}} = \frac{e}{n_{\rm e} \Omega} \sum_{n\in {\rm CB}}
  \int \frac{d\mathbf{k}}{\Omega_{\rm BZ}}  \frac{\partial f^0_{n\mathbf{k}}}{\partial 
  \varepsilon_{n\mathbf{k}}}  v_{n\mathbf{k},\alpha}  v_{n\mathbf{k},\beta} \tau_{n\mathbf{k}}. 
\end{equation}
The main advantage of the SERTA is that the grids of $\mathbf{k}$-points and $\mathbf{q}$-points do not need to be commensurate.
In practice, this allows for a denser sampling of the momentum regions that contribute the most to the mobility, i.e. regions close to the band edges. 
These grids typically converge faster than homogeneous or random grids for the same number of points. 
We now detail our computational setup used to evaluate Eq.~\eqref{eq.1}.

\begin{figure*}[t]
  \centering
  \includegraphics[width=0.9\linewidth]{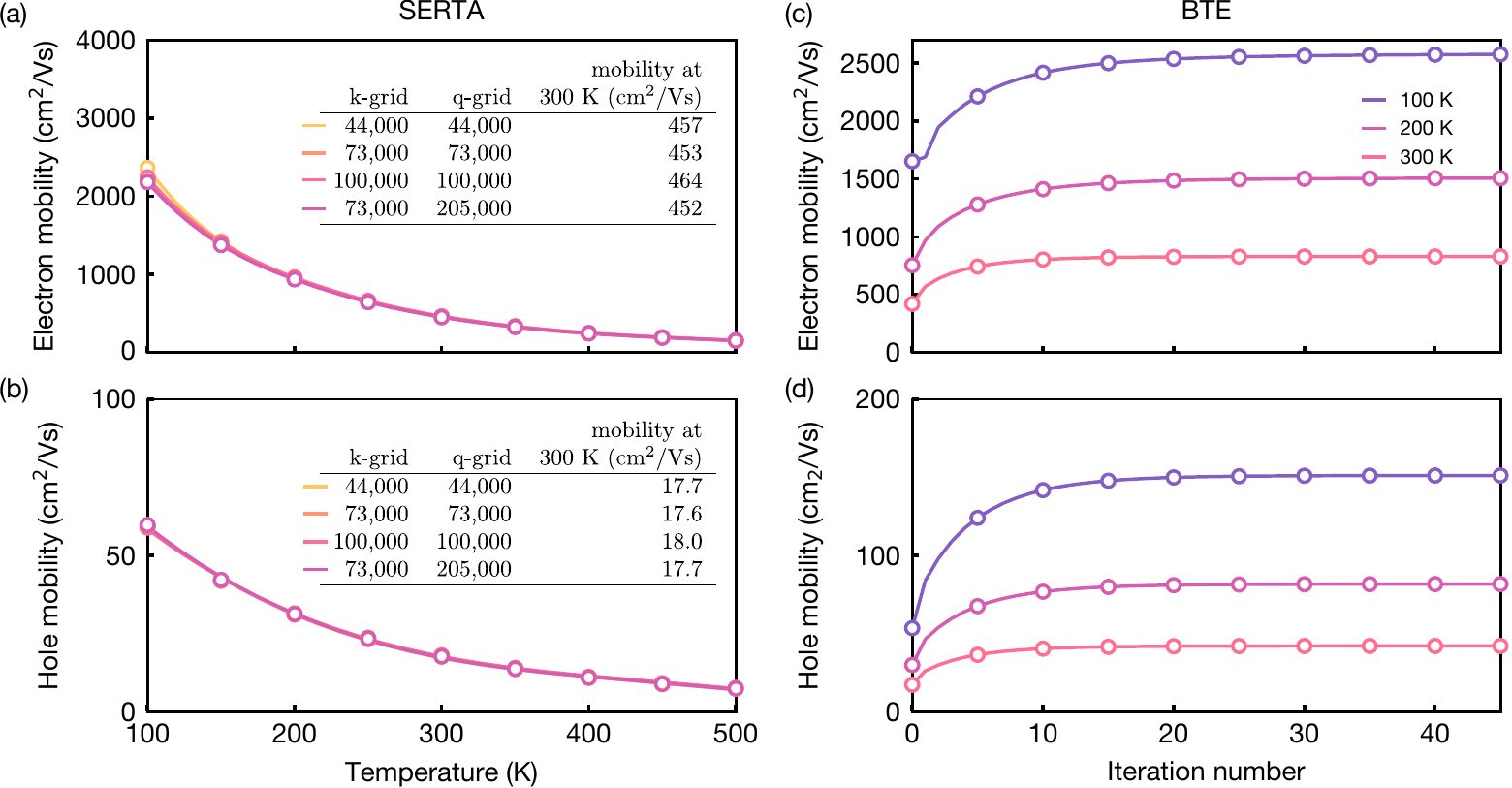}
  \caption{\label{figP4}
  Convergence tests for the electron and hole mobilities of wurtzite GaN. 
  (a-b) Electron and hole mobility vs.\ temperature, for different sizes of the Brillouin zone grids of electrons
  ($\bf k$) and phonons ($\bf q$). 
  The calculations are performed within the SERTA using LDA and Cauchy grids. 
  (c-d)~Electron and hole mobilities of GaN using uniform 100$\times$100$\times$100 grids, calculated
  iteratively from the self-consistent BTE, as a function of the number of iterations. 
  }
\end{figure*}

\subsection{Electron-phonon matrix elements and Brillouin-zone integrals}

\begin{figure*}[th]
  \centering
  \includegraphics[width=0.92\linewidth]{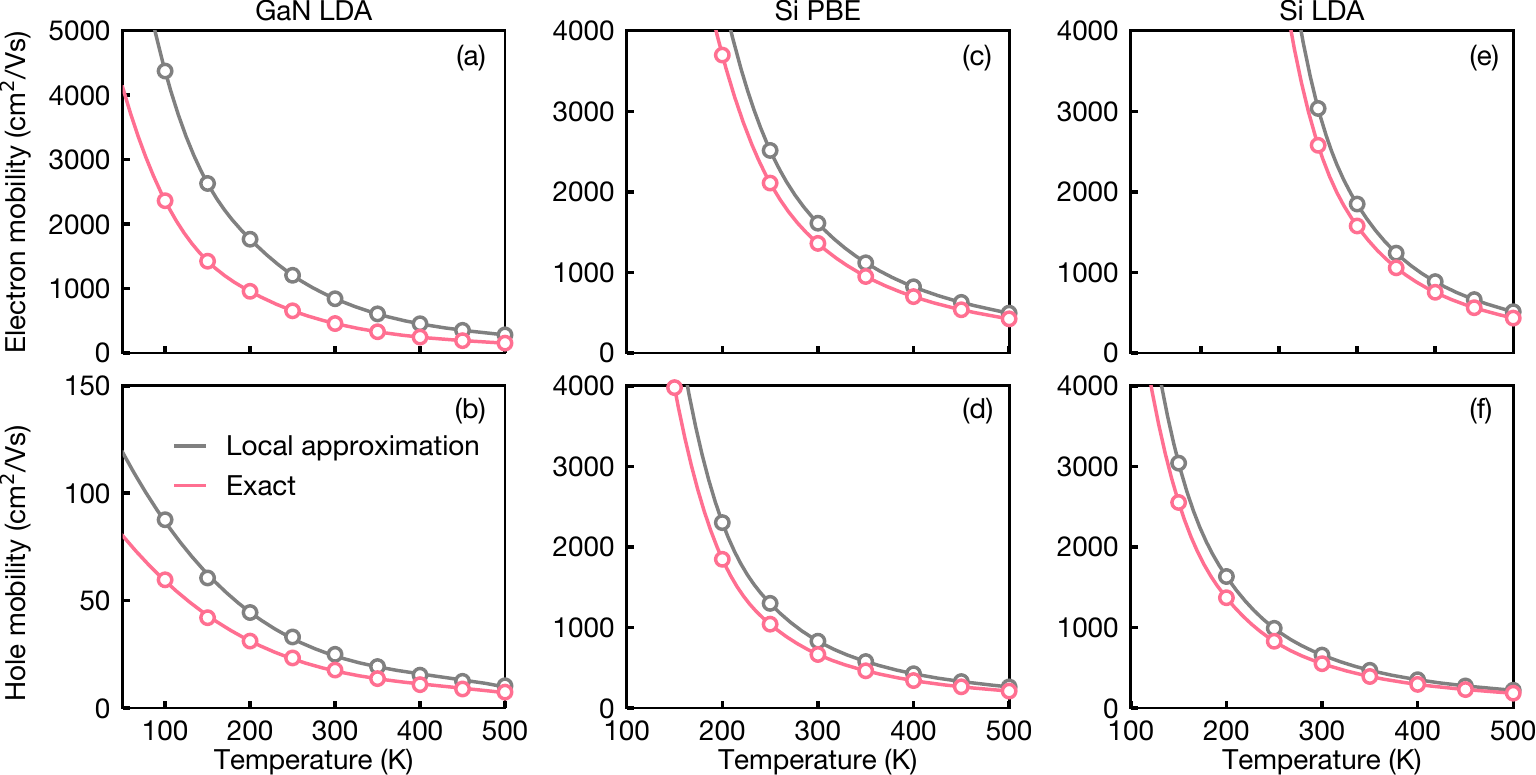}
  \caption{\label{figP5}
  Comparison between calculations of mobility using the ``local approximation'' to the band velocity
  and the ``exact'' velocity, which takes into account the contribution from the non-local part of the
  ionic pseudopotentials. (a-b) Electron and hole mobility in GaN versus temperature, both in the
  local approximation (grey) and using exact velocities (red). 
  Same comparison, this time for silicon, using a PBE pseudopotential (c-d) or LDA pseudopotential (e-f). 
  }
\end{figure*}

The key challenge in numerically evaluating Eq.~\eqref{eq.1} is related to the fact that the mobility converges very 
slowly with the number of $\mathbf{k}$ and $\mathbf{q}$-points included~\cite{Ponce2018}.
This translates into having to compute the electron-phonon matrix elements $g_{mn\nu}(\mathbf{k},\mathbf{q})$ from Eq.~\eqref{gmat} for millions 
of momentum points. 

In the case of SERTA calculations, we interpolate the electron-phonon matrix elements using Wannier 
functions~\cite{Giustino2007}, from a coarse 6$\times$6$\times$6 Brillouin-zone grid to a dense grid 
with 73,000~$\mathbf{k}$-points and 205,000~$\mathbf{q}$-points. The ${\bf q}$-points follow a Cauchy 
distribution of width $0.02~\text{\AA}^{-1}$ centered at $\Gamma$, and are weighted according to their 
Voronoi volume~\cite{Rycroft2009}. 
All the mobility calculations in this work are based on density-functional theory including spin-orbit coupling (SOC) for the 
Kohn-Sham states, density-functional perturbation theory for phonons and electron-phonon matrix elements, 
and many-body perturbation theory for GW quasiparticle corrections, as implemented in the software 
packages Quantum Espresso~\cite{Giannozzi2017}, Yambo~\cite{Marini2009}, wannier90~\cite{Mostofi2014}, 
and EPW~\cite{Giustino2007,Ponce2016a}.
The convergence tests for the SERTA drift mobility are presented on Fig.~\ref{figP4}(a-b). We observe a fast convergence with sampling size.
We obtained a room temperature electron and hole moblity of 452~cm$^2$/Vs and 18~cm$^2$/Vs, respectively.

In the case of the complete self-consistent solution of Eqs.~\eqref{eq.1}-\eqref{eq.3}, 
we use a homogeneous grid with 100$\times$100$\times$100 $\mathbf{k}$-points and 
$\mathbf{q}$-points. In this latter case we rely on the crystal symmetry operations on the $\mathbf{k}$-point grid to reduced the 
number of electron-phonon matrix elements to be explicitly computed.
We emphasize that, while in some systems the SERTA is accurate enough for predictive calculations 
of carrier mobilities~\cite{Fiorentini2016,Ponce2018}, this is not true in general~\cite{Ma2018}.
In the present case of wurtzite GaN, we find that the self-consistent solution of the BTE yields
enhancement factors of about 2 of the electron and hole mobility upon the values obtained within the SERTA,
as shown on Fig.~\ref{figP4}(c-d). 
This leads to electron and hole room temperature drift mobilities of 830~cm$^2$/Vs and 42~cm$^2$/Vs, respectively. 
Therefore it is very important to always benchmark SERTA results versus the complete solution of the BTE. 
The corresponding Hall mobility of 1034~cm$^2$/Vs and 52~cm$^2$/Vs for electron and hole, respectively; will be discussed in Sec.~\ref{mob_unstr_gan}.

We found that the self-consistent calculations converge rapidly, in about 20 iterations, and without the need for linear mixing. 
The convergence could be accelerated by methods such as conjugate gradients~\cite{Fiorentini2016},
but since the iterations are fast in comparison to the calculation of scattering rates, 
we find it unnecessary to improve the iterative solver.


\subsection{Band velocity}

In Ref.~\onlinecite{Ponce2018} the band velocities $v_{n{\bf k},\alpha}$ appearing in Eqs.~\eqref{eq.1} and \eqref{eq.2} for silicon were computed by neglecting the $\mathbf{k}$-derivatives 
of the ionic pseudopotentials. We named this approach the ``local velocity approximation'':
\begin{equation}\label{oldvelo}
v_{nm\mathbf{k},\alpha} =   \mathbf{k}\delta_{mn}  + \sum_{\mathbf{G}} c_{n\mathbf{k}}(\mathbf{G})^*c_{m\mathbf{k}}(\mathbf{G})\mathbf{G},
\end{equation}
where $c_{n\mathbf{k}}$ are the plane-waves coefficients. 
The ``exact'' band velocity can be computed as~\cite{Wang2006,Yates2007}:
\begin{equation}\label{newvelo}
v_{nm\mathbf{k},\alpha} = \frac{1}{\hbar} H_{nm\mathbf{k},\alpha} - \frac{i}{\hbar}(\varepsilon_{m\mathbf{k}}-\varepsilon_{n\mathbf{k}}) A_{mn\mathbf{k},\alpha},
\end{equation}
where $H_{nm\mathbf{k},\alpha}$ and $A_{mn\mathbf{k},\alpha}=i\langle u_{n\mathbf{k}}|\partial_\alpha u_{m\mathbf{k}}\rangle$ are the $\mathbf{k}$-derivatives of the Hamiltonian and position operator in the direction $\alpha$, interpolated on the fine momentum grids, and $u_{n\mathbf{k}}$ is the periodic part of the wavefunction.  
In both cases, the velocity $v_{n\mathbf{k},\alpha}$ is obtained by taking the diagonal elements of $v_{nn\mathbf{k},\alpha}$ from Eqs.~\eqref{oldvelo} or \eqref{newvelo}.

We find that the local velocity approximation is inadequate in GaN, and it can lead to an overestimation of the mobilities by up to 50\%. 
More specifically, the room temperature electron and hole mobility increases by 46\% and 30\%, respectively, when using Eq.~\eqref{newvelo} instead of Eq.~\eqref{oldvelo}. 
In the case of silicon and the LDA, the electron and hole mobility increase by 15\% and 16\%, respectively when using Eq.~\eqref{newvelo} instead of Eq.~\eqref{oldvelo}. 
As shown in Fig.~\ref{figP5}, this effect is not sensitive to the choice of the pseudopotential, but it depends strongly on the system under consideration.
Throughout this manuscript, we use the velocities given by Eq.~\eqref{newvelo}.

\begin{figure}[t!]
  \centering
  \includegraphics[width=0.92\linewidth]{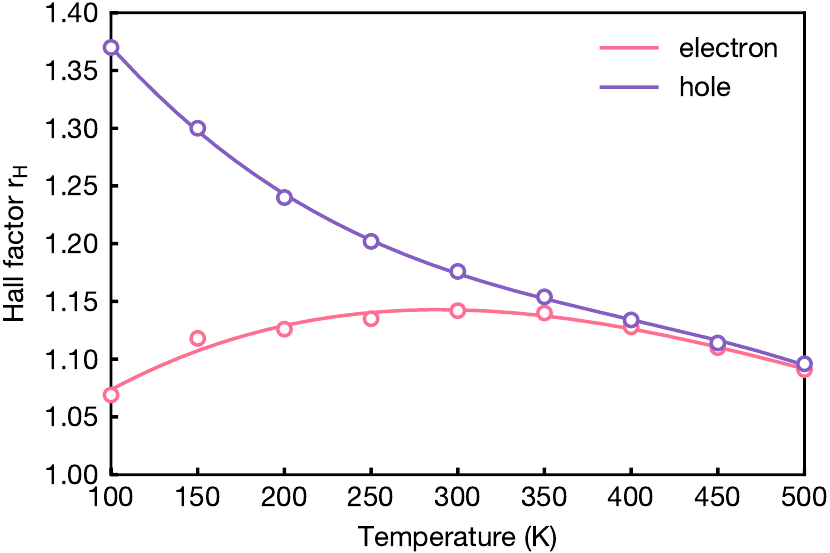}
  \caption{\label{figP6}
  Calculated temperature-dependent Hall factor of unstrained wurtzite GaN, for electrons and holes.
  }
\end{figure}

\subsection{Hall factor}

In many experiments it is common to measure the Hall mobility $\mu_{\rm H}$ instead of the drift mobility $\mu$ of Eq.~\eqref{eq.1}.
In order to perform meaningful comparisons, we calculate the Hall factor $r_{\rm H}$ and obtain the Hall mobilities $\mu_{\rm H} = r_{\rm H}\mu$. 
Following Ref.~\onlinecite{Wiley1975}, p.~118 and Ref.~\onlinecite{Price1957}, Eq.~(3.12), we calculate the temperature-dependent Hall factor 
as the ratio:
\begin{equation}\label{eq.4}
r_{\text{H}} = \langle \tau^2 \rangle/ \langle \tau \rangle^2,
\end{equation}
where 
\begin{equation}\label{eq.5}
\langle \tau^n \rangle 
= \frac{\int_0^\infty \tau^n(x) x^{3/2} e^{-x} dx}{\int_0^\infty x^{3/2} e^{-x} dx}
\end{equation}
is an energy-averaged
carrier scattering rate, and $x=\varepsilon/k_{\rm B}T$. 
The energy-dependent scattering rates are obtained through: 
\begin{equation}
\tau(\varepsilon) = \sum_{n} \int \frac{d\mathbf{k}}{\Omega_{\rm BZ}} \delta(\varepsilon-\varepsilon_{n\mathbf{k}}) \tau_{n\mathbf{k}},
\end{equation}
where the Dirac deltas are evaluated using Gaussian of width 1~meV. 
The calculated Hall factors for electrons and holes as a function of temperature are reported in Fig.~\ref{figP6}. 
The values range from 1.07 to 1.37 across the whole temperature range.  
These data are for unstrained GaN. 
We checked that the Hall factor is not sensitive to strain for the other cases considered in this work.

\subsection{Carrier mobility in unstrained GaN}\label{mob_unstr_gan}

\begin{table}[b]
  \begin{tabular}{c r r r r r}
  \toprule 
              &\multicolumn{5}{c}{Electron mobility (cm$^2$/Vs)} \\
  Temperature & \multicolumn{2}{c}{Drift mobility} & \multicolumn{3}{c}{Hall mobility} \\
    (K)       & SERTA & BTE  & BTE  & \multicolumn{2}{c}{Experiments} \\
              &       &      &  +scaling    & \\
         100  &  2363 & 3686 & \textbf{3941} & 3332~\cite{Kyle2014} & 2202~\cite{Goetz1998}  \\
         200  &   958 & 1916 & \textbf{2157} & 2420~\cite{Kyle2014} & 1700~\cite{Goetz1998}\\                     
              &       &      &               &  540~\cite{Ilegems1973} & 540~\cite{Goetz1996} \\  
         300  &   457 &  905 & \textbf{1034} & 1265~\cite{Kyle2014} & 840~\cite{Goetz1998}\\      
              &       &      &               &  330~\cite{Ilegems1973} & 370~\cite{Goetz1996} \\          
         400  &   247 &  480 &  \textbf{541} & 400~\cite{Goetz1998} & 160~\cite{Ilegems1973}\\
              &       &      &               & 245~\cite{Goetz1996} \\           
         500  &   154 &  299 &  \textbf{326} & 250~\cite{Goetz1998} & 100~\cite{Ilegems1973} \\
              &       &      &               & 150~\cite{Goetz1996} \\ 
              \\
              &\multicolumn{4}{c}{Hole mobility (cm$^2$/Vs)} \\
         100  &   60  &  168 &  \textbf{230} & - \\  
         200  &   31  &   85 &  \textbf{105} & 83~\cite{Horita2017} \\
         300  &   18  &   44 &   \textbf{52} & 31~\cite{Horita2017} \\   
         400  &   11  &   25 &   \textbf{28} & 14~\cite{Horita2017} \\  
         500  &    7  &   15 &   \textbf{16} & - \\                                                                                        
  \botrule 
  \end{tabular}
  \caption{\label{table2}
  Electron and hole mobilities of wurtzite GaN, calculated using the \textit{ab initio} Boltzmann
  formalism in the self-energy relaxation time approximation (SERTA) and iterative form (BTE), compared with experiment.
  We show both the drift mobilities computed via Eqs.~\eqref{eq.1}-\eqref{eq.3} and the Hall 
  mobilities obtained by applying the Hall factor shown in Fig.~\ref{figP6} (bold).  
  }
\end{table}

Using Eq.~\eqref{eq.1}, we computed the drift and Hall mobilities for electrons and holes in
intrinsic GaN as a function of temperature.
In table~\ref{table2}, we compare our results with available experimental
data. 
We find that the Hall mobility is about 15\% higher than the drift mobility at room
temperature, as expected~\cite{Lundstrom2009}.
Our predicted electron and hole Hall mobilities at 300~K are 1034~cm$^2$/Vs and 52~cm$^2$/Vs, respectively. They are in good agreement with the measured values  
1265~cm$^2$/Vs~\cite{Kyle2014} and 31~cm$^2$/Vs~\cite{Horita2017}, respectively. 
%
The Hall mobility at 100~K is computed to be 3941~cm$^2$/Vs for electrons and 230~cm$^2$/Vs for holes. 
Since at room temperature the Cauchy grid yields mobilities which are converged within 1\%, and uniform grids yield mobilities converged 
within 10\%, we use the ratio between the BTE and SERTA mobilities on uniform grids to estimate the BTE mobilities on dense Cauchy grids. 
Direct BTE calculations are not possible on such grids due to the commensurability requirement in Eq.~\eqref{eq.2}. 
The results reported throughout the manuscript correspond to this ratio; the conclusions of the manuscript remain unchanged if we use the BTE results for homogeneous grids.

In table~\ref{table2}, we see that the electron mobility is $1034/52\approx 20$ times higher than the hole mobility. 
Experimentally, this ratio is even larger, $1265/31\approx 41$.
To understand the origin of the large difference between the electron and hole mobilities
in GaN, we refer to Eqs.~\eqref{eq.1}-\eqref{eq.3}.
In the simplified case of parabolic bands, the mobility in Eq.~\eqref{eq.1} scales as $e \tau/ m^*$ following Drude's law,
with $m^*$ and $\tau$ being the average effective mass and relaxation rate, respectively. 
%
As discussed in Section~\ref{GWcorr}, Table~\ref{table1} and Fig.~\ref{figP3}, 
the ratio between the conductivity effective masses $3/(m_{\parallel}^{-1}+2 m_{\perp}^{-1})$ of electrons and holes is 2.4/2.9 for the $hh$/$lh$ case, respectively. 
These values are significantly lower than the observed ratio of electron to hole mobilities, therefore the difference between electron 
and hole effective masses alone cannot fully account for the order-of-magnitude difference in carrier mobilities.

To determine the origin of the residual difference between electron and hole mobilities, we analyzed in Ref.~\onlinecite{Ponce2019} the angular averages of the carrier relaxation rate $1/\tau$. 
Although every electronic state has its own lifetime $\tau_{n\mathbf{k}}$ in our calculation, we have shown previously that the most 
representative carrier energy~\cite{Ponce2019a} comes out from an energy $k_{\rm B}T = 25$~meV away from the band edges, and these are the value that we used for our analysis in Ref.~\onlinecite{Ponce2019}.

By examining the scattering rates and their spectral decomposition $\partial \tau^{-1}/\partial\omega$ in terms of phonon energy, 
we found that the dominant scattering channel is from long-wavelength acoustic phonons around a phonon energy of 2~meV (77\% and 84\% of the scattering rates for electrons and holes, respectively).
The remaining contribution is from polar Fr\"ohlich longitudinal-optical (LO) phonons near $\Gamma$, and located around a phonon energy of 91~meV both for electron and holes. 
The largest source of acoustic scattering in the case of holes is from acoustic-deformation-potential (ADP) scattering. 

In the case of ADP scattering, the scattering rate scales with the electronic density of states, and hence with the effective masses, 
as $1/\tau \sim (m^{*})^{3/2}$~\cite{Lundstrom2009}.
Using the angular averages, we computed the electron lifetimes to be in the range of 17~fs, while the hole lifetimes of around 4~fs are much shorter. 
%
Their ratio is similar to the ratio between the conductivity effective masses, $(m_{lh}^{*})^{3/2}/(m_{\rm e}^{*})^{3/2} = 3.7$ and $(m_{hh}^{*})^{3/2}/(m_{\rm e}^{*})^{3/2} = 4.9$. 
This highlights the fact that the high density of $lh$ and $hh$ states plays a central role in reducing the hole mobility.

The combination of higher effective masses and larger density of states account for most of the mobility difference between electron and hole in GaN. 
The remaining difference can be attributed to other effects that are also responsible for reducing hole mobility.
For example the strong non-parabolicity of the $hh$ in-plane band will increase the effective masse for states with momentum sligthly away from the zone center~\cite{Kim1996,Rinke2008};
the fact that GaN has multiple scattering channels for the holes (two spin-split sets of bands) will also increase the overall scattering~\cite{Yeo1998,Svane2010}; and
longitudinal-optical phonons contribute about 20\% of additional scattering. 
All these effects are fully accounted for in our \textit{ab initio} BTE formalism.

Now that we have a better understanding of the various mechanisms behind the low hole mobility, we can
proceed to computational design of higher-mobility p-type GaN.
Since the low mobilities stem primarily from the presence of two adjacent bands with heavy masses, 
we investigate whether we can employ strain to change the energetics and ordering of the valence band maximum states.

\section{Carrier mobility of strained GaN}\label{carr_mob_strain}

\subsection{Elastic properties}\label{elastic_cst}

We studied the elastic properties of wurtzite GaN using the thermo\_pw code~\cite{Corso2016}.
The stiffness matrix $C_{ij}$ was obtained by third-order polynomial fitting under 12 deformations geometries of small strain intervals of 0.001 to remain in the linear regime.
For each strain, the ions were relaxed to their equilibrium positions.  
The stiffness matrix of Laue class $D_{6h}$ for wurtzite crystals has five independent elastic constants $C_{11}$, $C_{12}$, $C_{13}$, $C_{33}$ and, $C_{44}$. 
The other coefficients follow the symmetry relationships $C_{23}=C_{13}$, $C_{55}=C_{44}$ and $C_{66}=(C_{11}-C_{12})/2$. 
%
%
The computed stiffness constants are given in table~\ref{table3} and are compared to prior theoretical and experimental values. 
In the Voigt approximation, the bulk and shear moduli are given by~\cite{Hill1952}: 
\begin{align}\label{eq:Bv}
9B_{\rm V}     =& C_{11}+C_{22}+C_{33} + 2(C_{12}+C_{13}+C_{23}),\\
15 G_{\rm V}   =& C_{11}+C_{22}+ C_{33} - (C_{12}+C_{13}+C_{23})   \nonumber \\
                                 &+ 3 (C_{44} + C_{55} + C_{66}), \label{eq:Br}
\end{align}
while in the Reuss approximation, the bulk and shear modulus are defined as~\cite{Hill1952}: 
\begin{align}
B_{\rm R}^{-1} =& S_{11}+S_{22}+S_{33} + 2(S_{12}+S_{13}+S_{23}), \label{eq:Gv} \\
15 G_{\rm R}^{-1} =& 4(S_{11}+S_{22}+ S_{33}) - 4(S_{12}+S_{13}+S_{23}) \nonumber \\
                                   &+ 3(S_{44} + S_{55} + S_{66}),\label{eq:Gr}
\end{align}
where $S_{ij} = C_{ij}^{-1}$ is the elastic compliance matrix. 
The Voigt approximation provides an upper bound for the bulk and shear moduli, while the Reuss approximation gives a lower bound. 
We can therefore define the arithmetic mean, refered to as the Void-Reuss-Hill approximation~\cite{Hill1952}, as $B=(B_{\rm V}+B_{\rm R})/2$ and 
$G=(G_{\rm V}+G_{\rm R})/2$. 
We then express effective Young $E$ modulus and Poisson ratio $\nu$ as:
\begin{align}
E   =& 9B G/(3B+G) \label{eq:E}, \\
\nu =& (3B-2G)/(6B+2G). \label{eq:nu}
\end{align}
Using Eqs.~\eqref{eq:Bv}-\eqref{eq:nu}, we obtained a bulk modulus of 202~GPa, a Young modulus of 358~GPa, a shear modulus of 148~GPa, and a Poisson's ratio of 0.205.
As expected with the LDA, our calculations overestimate the experimental values of 188~GPa~\cite{Xia1993}, 295~GPa~\cite{Nowak1999}, and 116~GPa~\cite{Yonenaga2005} for the bulk, Young and shear modulus, respectively. 
In contrast, the calculated Poisson's ratio sits in between the experimental values 0.183~\cite{Moram2007} and 0.23~\cite{Kisielowski1996}.
%

\subsection{Phonon dispersion relations}\label{strain_phBS}

\begin{table}[th]
  \begin{tabular}{l c c c c c c }
  \toprule 
                         & $C_{11}$ & $C_{12}$ & $C_{13}$ & $C_{33}$ & $C_{44}$ & $C_{66}$ \\
\rule{0pt}{1em}Present work &  (GPa)   &   (GPa)       &    (GPa)      &     (GPa)     &     (GPa)     &     (GPa) \\
\hline                       
LDA                      & 435 & 108 &  65 & 474 & 115 & 163  \\
\rule{0pt}{1em}Previous work&  &     &     &     &     &      \\
\hline 
LDA~\cite{Kim1997a}      & 346 & 148 & 105 & 405 & 76  &  99  \\
LDA~\cite{Wright1997}    & 367 & 135 & 103 & 405 & 95  & 116  \\
LDA~\cite{Wagner2002}    & -   &   - & 104 & 414 & -   & -    \\ 
LDA~\cite{Qin2017}       & 374 & 127 &  81 & 442 &  99 & 124  \\
Exp.~\cite{Polian1996}   & 390 & 145 & 106 & 398 & 105 & 123  \\
Exp.~\cite{Yamaguchi1997}& 365 & 135 & 114 & 381 & 109 & 115  \\
Exp.~\cite{Deger1998}    & 370 & 145 & 110 & 390 &  90 & 112  \\ 
Exp.~\cite{Deguchi1999}  & 373 & 141 & 80  & 387 &  94 & 118  \\  
\\
    
                              &  & $B$     &  $E$     & $G$      & $\nu$ & \\
\rule{0pt}{1em}Present work   &  & (GPa) &  (GPa) & (GPa)  &       & \\
\hline                           
LDA                           &  &   202 &   358  &  148   & 0.205 & \\
\rule{0pt}{1em}Previous work  &  &       &        &        &       & \\
\hline 
LDA~\cite{Kim1996}            &  &   207 &    - &    -   & - & \\
LDA~\cite{Wright1997}         &  &   202 &    - &    -   & - & \\
LDA~\cite{Wagner2002}         &  &   207 &    373 &    -   & 0.202 & \\
LDA~\cite{Qin2017}            &  &   196 &    303 &  122   & 0.240 & \\ 
Exp.~\cite{Xia1993}           &  &   188 &      - &    -   & -     & \\
Exp.~\cite{Polian1996}        &  &   210 &    356 &  120   & 0.198 & \\ 
Exp.~\cite{Yamaguchi1997}     &  &   205 &    293 &  116   & 0.261 & \\ 
Exp.~\cite{Deger1998}         &  &   207 &    276 &  108   & 0.278 & \\                                                                 
Exp.~\cite{Deguchi1999}       &  &   192 &    286 &  114   & 0.252 &\\ 
Exp.~\cite{Nowak1999}         &  &     - &    295 &    -   & 0.250 & \\
Exp.~\cite{Yonenaga2005}      &  &     - &    295 &  116   & 0.250 & \\                                                                     
  \botrule 
  \end{tabular}
  \caption{\label{table3}
  Comparison between calculated elastic constants $C_{ij}$, bulk $B$, Young $E$, shear $G$ modulus, and Poisson's ratio $\nu$ (LDA calculations without spin-orbit coupling) with prior theoretical and experimental work. 
  }
\end{table}

\begin{figure*}[th]
  \centering
  \includegraphics[width=0.90\linewidth]{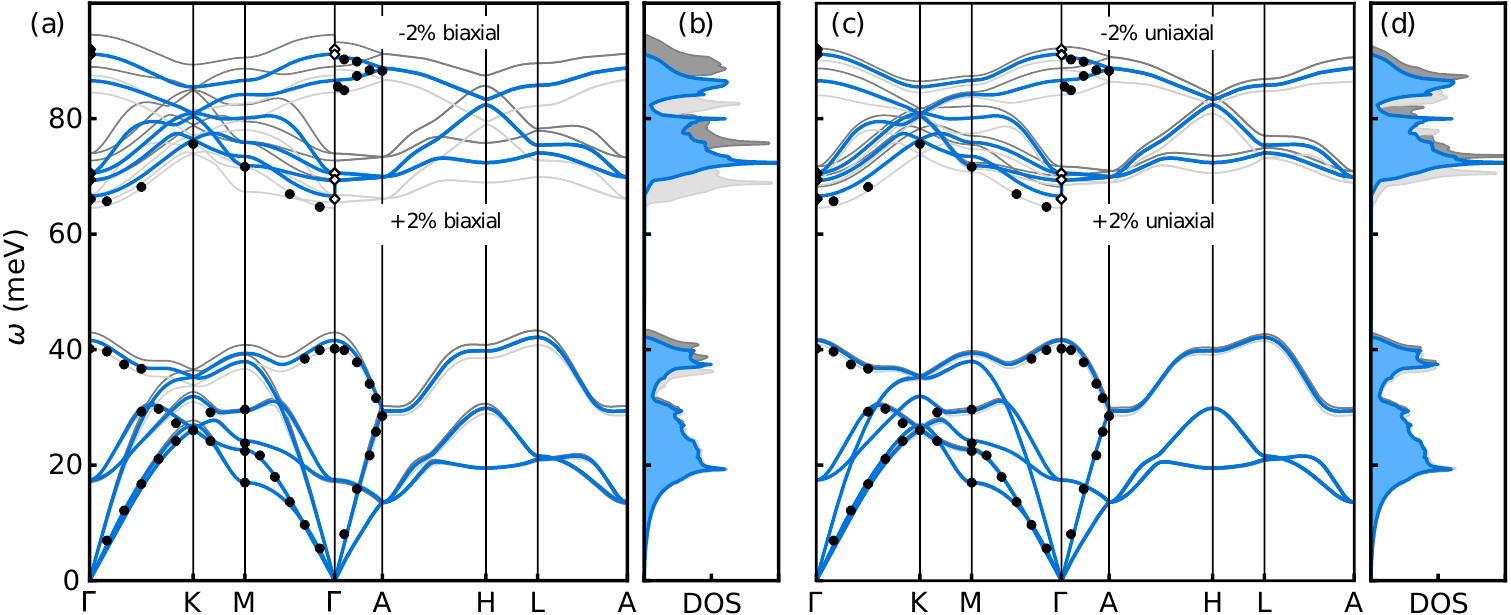}
  \caption{\label{figP7}
   Phonon dispersion relations (a,c) and phonon density of states (b,d) of wurtzite GaN at the relaxed (blue lines) or strained (gray lines) atomic positions.
   The experimental data are from Ref.~\onlinecite{Ruf2001} (filled discs, inelastic X-ray scattering) and from
  Ref.~\onlinecite{Siegle1997} (empty diamonds, Raman). 
  }
\end{figure*}

We compute the phonon dispersions and phonon density of states using density functional perturbation theory~\cite{Gonze1997a,Baroni2001} as implemented in Quantum Espresso~\cite{Giannozzi2017}, for unstrained GaN as well as under $\pm 1\%$ and $\pm 2\%$ biaxial and uniaxial strain.  
The phonon dispersions of the unstrained GaN are reported in Fig.~\ref{figP7} in blue.
The theoretical curves follow closely the experimental data from inelastic X-ray scattering~\cite{Ruf2001} and the Raman measurements~\cite{Siegle1997}.
The calculated unstrained in-plane ($\varepsilon_{\infty}^{\perp}$) and out-of-plane ($\varepsilon_{\infty}^{\parallel}$) high-frequency dielectric constants are 5.60 and 5.77, respectively. 
These values slightly overestimate the experimental values of $\varepsilon_{\infty}^{\perp}$=5.14~\cite{Yu1997}, 5.25~\cite{Hibberd2016},  5.29~\cite{Azuhata1995} or 5.35~\cite{Barker1973} as well as $\varepsilon_{\infty}^{\parallel}$=5.31~\cite{Yu1997}, 5.41~\cite{Hibberd2016}.
The corresponding in-plane and out-of-plane Born effective charges are 2.59 and 2.73, respectively; these values are in agreement with earlier theoretical values of 2.60 and 2.74, respectively~\cite{Wagner2002}.

We now investigate how the phonon dispersions change under strain. 
As shown in Fig.~\ref{figP7}(a), the zone-center highest E$_1$ LO phonon hardens from 91.2~meV to 94.5~meV under 2\% biaxial compressive strain, and softens to 87.5~meV under 2\% tensile biaxial strain.
The same behavior occurs under uniaxial strain, although changes are more modest, as seen in Fig.~\ref{figP7}(b): the highest phonon mode hardens to 92.2~meV under 2\% compression, and softens to 90.1~meV under 2\% traction. 
This behavior can easily be understood by the fact that a fixed uniaxial strain imposed to a crystal has a smaller effect than a corresponding fixed biaxial strain. Indeed a $\pm$ 2\% uniaxial strain modifies the overall volume from 98.8\% to 101.2\% while for biaxial strain, the change of volumes goes from 97.0\% to 103.0\%.

The high-frequency dielectric constants, $\varepsilon_{\infty}^{\perp}$ and $\varepsilon_{\infty}^{\parallel}$ are almost insensitive to
uniaxial and biaxial strain, respectively; 
while $\varepsilon_{\infty}^{\parallel}$ increases from 5.62 (-2\% strain) to 5.96 (+2\% strain) under uniaxials strain, and 
 $\varepsilon_{\infty}^{\perp}$ increases from 5.44 (-2\% strain) to 5.80 (+2\% strain) under biaxial strain.
Finally, the Ga and N Born effective charges have opposite values and the largest change is observed for in-plane biaxial strain, going from 2.56 (-2\% strain) to 2.62 (+2\% strain).

\subsection{Band structures of strained GaN}\label{strain_elBS}

\begin{figure*}[ht]
  \centering
  \includegraphics[width=0.94\linewidth]{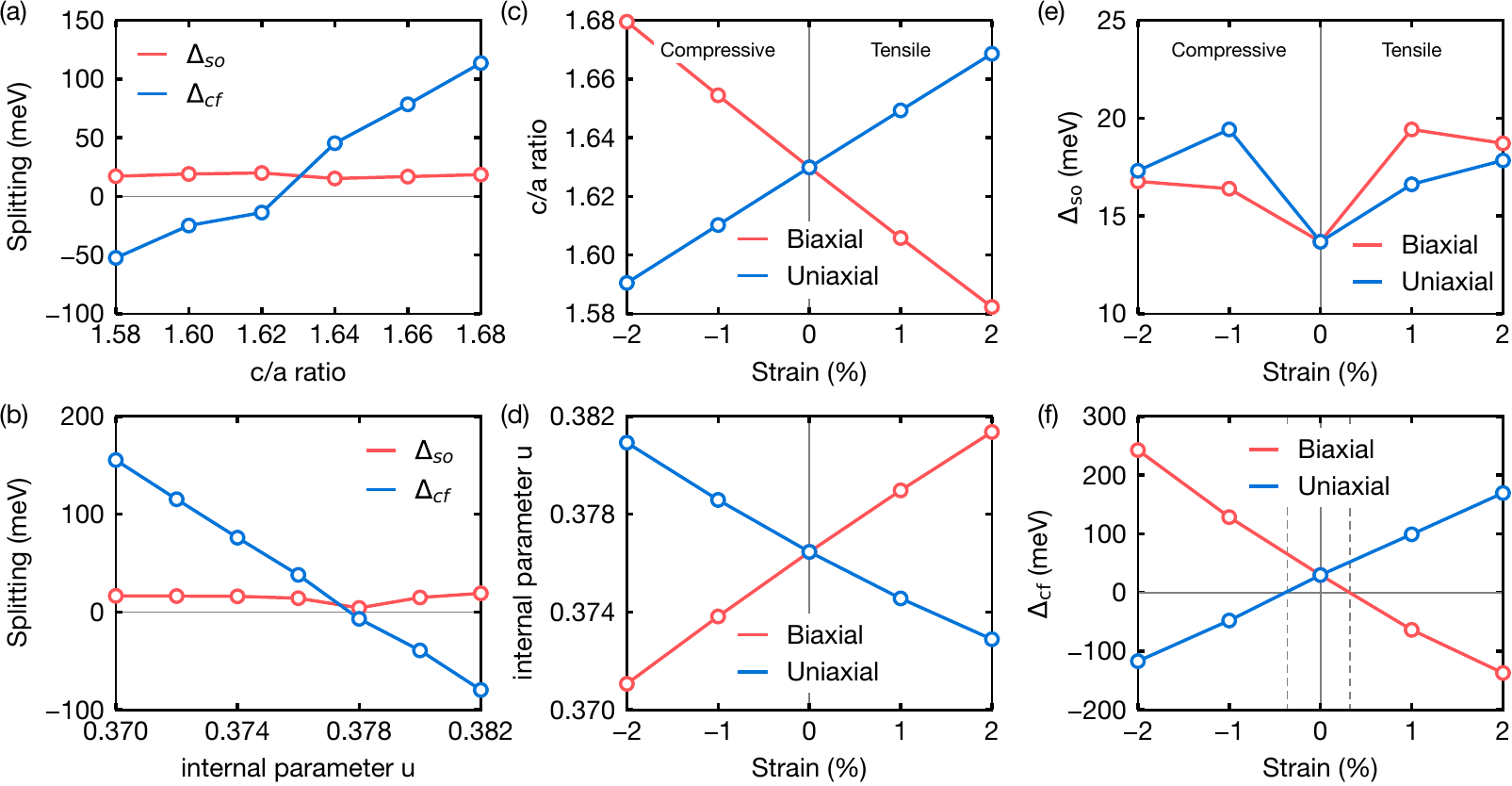}
  \caption{\label{figP8}
  (a-b) Sensitivity of the spin-orbit splitting $\Delta_{\rm so}$ and the crystal-field splitting 
  $\Delta_{\rm cf}$ to the $c/a$ ratio and the internal $u$ parameter in wurtzite GaN.
  In (a) we fix the internal parameter to $u=0.3765$, in (b) we fix the aspect ratio
  to $c/a=1.6299$. 
  (c-d) Dependence of $c/a$ and $u$ on biaxial and uniaxial strain, respectively.
  (e-f) Dependence of $\Delta_{\rm so}$ and $\Delta_{\rm cf}$ on the biaxial and uniaxial strain, respectively. 
  The region outside of the dashed lines in (f) leads to a reversal of $\Delta_{\rm cf}$.
  }
\end{figure*}

We now turn to the changes of electronic properties under uniaxial and biaxial strain. 
As seen in Eq.~\eqref{delat_so}, the $lh$ and $hh$ bands are separated by the spin-orbit splitting $\Delta_\text{so}$,
which we found to be relatively insensitive to the $c/a$ ratio and internal parameter $u$, see Fig.~\ref{figP8}(a-b).
However, the separation between the $lh$/$hh$ and the $sh$ bands is controlled by the crystal-field splitting, $\Delta_{\text{cf}}$ which is given by Eq.~\eqref{delat_cf}.
In contrast to $\Delta_\text{so}$, the crystal-field splitting is known to be sensitive to the internal parameter $u$ of the wurtzite structure, 
or equivalently to a change of the $c/a$ ratio~\cite{Kim1996,Yan2009}.
As seen in Fig.~\ref{figP8}(a-b), $\Delta_{\text{cf}}$ increases from -52~meV to +114~meV following the increase of the $c/a$ ratio from 1.58 to 1.68. 
In contrast, the crystal-field splitting decreases linearly from 155~meV to -80~meV as the internal parameter $u$ increases from 0.370 to 0.382.
In Fig.~\ref{figP8}(c-d), we can see that there is a linear correlation between the applied strain and the $c/a$ or internal parameter $u$.
In the case of biaxial strain, the $c/a$ ratio decreases with strain while $u$ increases and the situation is reversed for uniaxial strain. 

We therefore see how by combining Fig.~\ref{figP8}(a-b) and (c-d) we can modify $\Delta_{\rm so}$ and $\Delta_{\rm cf}$ via strain. 
In particular, we see in Fig.~\ref{figP8}(e) that the $\Delta_{\rm so}$ has a minimum for unstrained GaN, and slightly increases with strain; while
in Fig.~\ref{figP8}(f) we see a drastic linear reduction of $\Delta_{\rm cf}$ with increasing biaxial strain, going from +243~meV to -137~meV.
A smaller linear increase of $\Delta_{\rm cf}$ is observed with increasing uniaxial strain, from -117~meV to +169~meV for -2\% and +2\% strain, respectively. 
We emphasize that the reason for a smaller splitting in Fig.~\ref{figP8}(a-b) is that these results are obtained at fixed $u$ or $c/a$ ratio, while in Fig.~\ref{figP8}(e-f) both $c/a$ and $u$ are changing with strain, contributing to increase the splitting even more. 
We see that in both types of applied strain, a reversal of the crystal field splitting is observed. 
Such reversal of the crystal-field splitting happens for biaxial tensile strain (along 
[2$\overline{1}\overline{1}$0] and [$\overline{1}$2$\overline{1}$0]) and for uniaxial compressive strain (along [0001]).
Using linear interpolation of our results from Fig.~\ref{figP8}(f), we estimate that the  crystal-field splitting reversal will happen 
at +0.46\% biaxial tensile strain and -0.62\% uniaxial compressive strain.  
\begin{figure}[b!]
  \centering
  \includegraphics[width=0.99\linewidth]{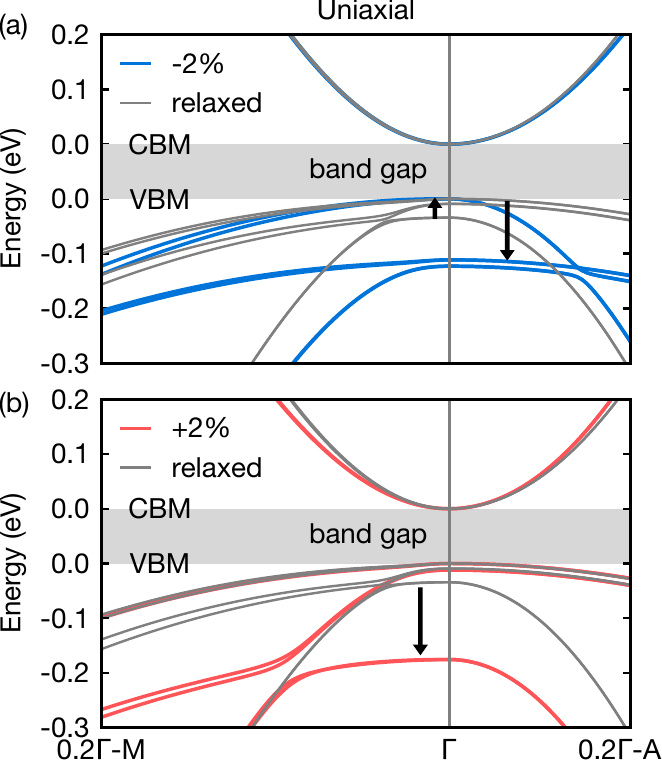}
  \caption{\label{figP9}
  (a-b) Change in the GW quasiparticle band structure of GaN upon uniaxial compression and dilation, respectively. 
  The energy levels have been aligned to the band edges.
  }
\end{figure}
Under these conditions, the split-off hole band is lifted above the light-hole and heavy-hole bands, as shown in Fig.~\ref{figP9}(a) for the case of uniaxial compression.

In contrast, in the case of applied uniaxial tensile strain shown in Fig.~\ref{figP9}(b), we can see that the $hh$ band is almost unaffected with respect to the unstrained cases, while the $sh$ and $lh$ bands get significantly pushed down in energy.
The largest change of band energy with strain are indicated with black arrows in Fig.~\ref{figP9}.
A similar but reversed effect is observed with biaxial strain, as shown in Ref.~\onlinecite{Ponce2019}. 
In that case the $sh$ band goes above the $hh$ and $lh$ bands under biaxial tensile strain.
This effect alters the ordering of the valence band top, as well as the character of the wavefunctions.
As discussed in Ref.~\onlinecite{Ponce2019}, the hole wavefunction at the valence band maximum has a dominant N-$p_{x,y}$ character. 
In both uniaxial and biaxial strain, the hole wavefunction keeps this character if no band reversal occurs, but
as soon as the $sh$ bands goes above the $hh$ and $lh$ bands, it abruptly changes character to a dominant N-$p_z$ states.

As seen in Section~\ref{GWcorr}, the conductivity effective mass of the $sh$ band is $m^*_{sh} = 0.45\,m_{\rm e}$ at the 
zone center, but away from $\Gamma$ it quickly decreases to $m^*_{sh} = 0.22\,m_{\rm e}$ due to strong 
non-parabolicity.
This effective mass is much smaller than the masses of the light and heavy hole bands. 
For this reason, we expect the hole mobility of GaN to sharply increase upon reversal of the sign of $\Delta_\text{cf}$.
In the next section, we validate this assumption by performing first-principles calculations of mobility on strained GaN.

\subsection{Mobility of strained GaN}\label{strain_mob}

\begin{table*}
  \begin{tabular}{r c r r r@{\hskip 20pt}r r@{\hskip 20pt}r r r@{\hskip 20pt}r r}
  \toprule\\[-6pt]
               &                 &  \multicolumn{5}{c}{Holes}  & \multicolumn{5}{c}{Electrons}  \\[3pt]
               &                 &  \multicolumn{3}{c}{\!\!\!\!Homogenous grid} & \multicolumn{2}{c}{Cauchy grid}  & \multicolumn{3}{c}{\!\!\!\!Homogenous grid} & \multicolumn{2}{c}{Cauchy grid}  \\[3pt]
               &                 &  \multicolumn{2}{c}{Mobility } & & \multicolumn{2}{c}{Mobility} &  \multicolumn{2}{c}{Mobility } & & \multicolumn{2}{c}{Mobility} \\[2pt]
               &  Temperature    &  \multicolumn{2}{c}{(cm$^2$/Vs)} & & \multicolumn{2}{c}{(cm$^2$/Vs)} & \multicolumn{2}{c}{(cm$^2$/Vs)} & & \multicolumn{2}{c}{(cm$^2$/Vs)}\\[3pt]       
   Strain      & (K)             & \multicolumn{1}{r}{SERTA}& \multicolumn{1}{c}{BTE} & \multicolumn{1}{c}{\!\!\!Ratio} & \multicolumn{1}{r}{SERTA} & \multicolumn{1}{c}{BTE$^\dagger$} & \multicolumn{1}{r}{SERTA} & \multicolumn{1}{c}{BTE} & \multicolumn{1}{c}{\!\!\!Ratio} & \multicolumn{1}{r}{SERTA} & \multicolumn{1}{r}{BTE$^\dagger$} \\[5pt]
              &      & $a$  & $b$ &  $\,\,\,\,c=b/a$ & $d$ &\,\,\,\textbf{$e\!=\!c\!\cdot\! d$}& $f$ & $g$ & \,\,\,$h=g/f$ & $i$ &\,\,\,\textbf{
  $j\!=\!h\!\cdot\! i$}\\[3pt]
\multirow{2}{*}{2\% Biaxial}
              &  100 & 289  & 571 &  1.98 & 231 &\textbf{457}& 1995 & 3482 & 1.75 & 2466 &\textbf{ 4315}\\
              &  300 &  54  & 113 &  2.09 &  46 &\textbf{ 96}&  438 &  847 & 1.93 &  459 &\textbf{  886}\\[5pt]
\multirow{2}{*}{1\% Biaxial}
              &  100 & 226  & 478 &  2.11 & 177 &\textbf{373}& 1945 & 2836 & 1.46 & 2754 &\textbf{ 4021}\\
              &  300 &  38  &  84 &  2.21 &  32 &\textbf{ 71}&  453 &  835 & 1.84 &  506 &\textbf{  931}\\[5pt] 
\multirow{2}{*}{Unstrained}
              &  100 &  54  & 151 &  2.80 &  60 &\textbf{168}& 1652 & 2584 & 1.56 & 2363 &\textbf{ 3686}\\ 
              &  300 &  17  &  42 &  2.47 &  18 &\textbf{ 44}&  420 &  830 & 1.98 &  457 &\textbf{  905}\\[5pt]
\multirow{2}{*}{-1\% Uniaxial}
              &  100 & 176  & 379 &  2.15 & 139 &\textbf{299}& 1567 & 2243 & 1.43 & 2178 &\textbf{ 3114}\\
              &  300 &  30  &  67 &  2.23 &  25 &\textbf{ 56}&  404 &  730 & 1.81 &  439 &\textbf{  795}\\[5pt]
\multirow{2}{*}{-2\% Uniaxial}
              &  100 & 249  & 519 &  2.08 & 206 &\textbf{428}& 2014 & 2350 & 1.17 & 1976 &\textbf{ 2312}\\
              &  300 &  53  & 117 &  2.21 &  46 &\textbf{102}&  431 &  724 & 1.68 &  408 &\textbf{  685}\\[3pt]                              
  \botrule 
  \end{tabular}
  \caption{\label{table4}
  Calculated drift mobilities of GaN for several strain levels and temperatures within the SERTA approximation and the more accurate BTE 
  using uniform grids, SERTA results using Cauchy grids, 
  and BTE results estimated from these data (``BTE$^\dagger$'', boldface).
  We use the ratio between the BTE ($b$) and SERTA ($a$) mobilities on uniform grids to estimate
  the BTE mobilities on Cauchy grids, $e=(b/a)\cdot c$. 
  }
\end{table*}

We performed transport calculations for uniaxially and biaxially strained GaN.
We computed the drift mobilities of GaN for several strain levels and temperatures using the SERTA approximation with Cauchy grids, and using the more accurate BTE with uniform grids. 
The uniform grids consist of $100\times 100\times 100$ $\mathbf{k}$- and $\mathbf{q}$-points while the random Cauchy grids consist of 45,000 points.
The calculated mobilities at different temperature within the SERTA or BTE are presented in Table~\ref{table4}.
First we computed the mobilities on the homogenous grid within the SERTA and the iterative solution. 
We observed an increase in hole mobility ranging between a factor 1.98 to 2.8 across the entire strain and temperature ranges when using the BTE compared to the SERTA. 
For electrons the ratio is slightly more modest, ranging from 1.17 to 1.98.

As expected, the room-temperature hole mobility significantly increases from 42~cm$^2$/Vs to 113 (117)~cm$^2$/Vs upon 2\% biaxial tensile (uniaxial compressive) strain. 
Conversely, the hole mobility remains almost constant when the crystal experiences no reversal of the crystal field splitting, i.e. under compressive biaxial or tensile uniaxial strain.

As discussed in Section~\ref{mob_unstr_gan}, Cauchy grids converge faster than homogeneous grids, but do not allow for BTE calculations due to incommensurablilty of the momentum grids. 
However, we noted that the ratio between the BTE to SERTA results is converging faster than the value themselves. 
Therefore we used the ratio between the BTE and SERTA mobilities on uniform grids to estimate the BTE mobilities on Cauchy grids. The results are shown in Table~\ref{table4} in bold, and represent our most accurate estimates.

In contrast to the hole mobility, the electron mobility remains close to the value for unstrained GaN  in the case of biaxial strain, but decreases slightly in the case of uniaxial strain due to a small increase in the electron effective mass. 
%
%
For this reason, the use of biaxial strain might be more attractive than uniaxial strain, as it leads to a doubling of the hole mobility with no change to the electron mobility. 
However, given that the electron mobility is already high and the reduction under strain is in the order of 20\%, one could easily imagine a successful device based on uniaxially-strained GaN.

\begin{figure}[h]
  \centering
  \includegraphics[width=0.90\linewidth]{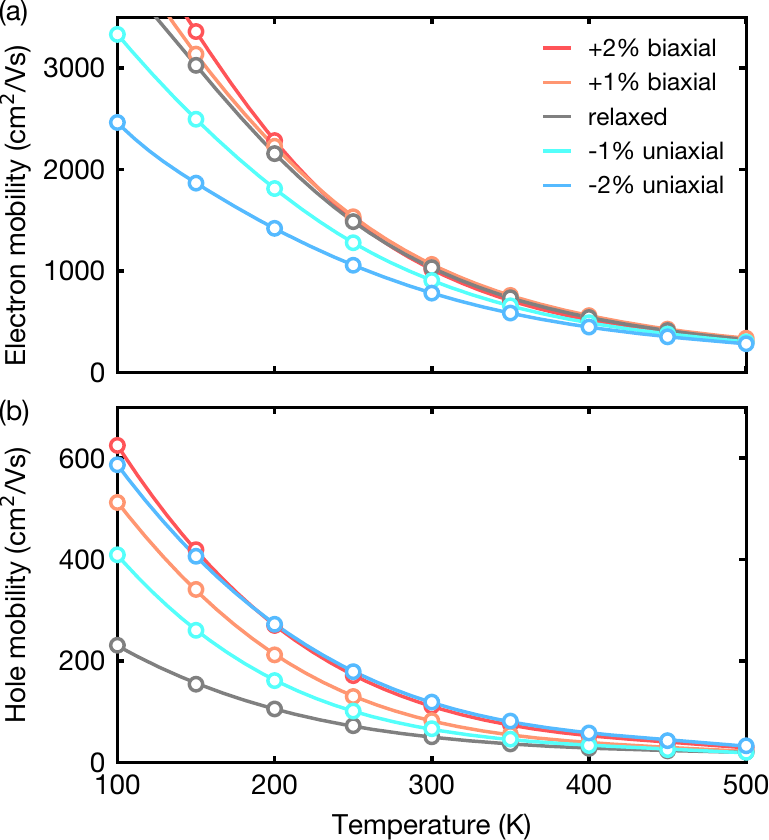}
  \caption{\label{figP10}
Predicted temperature-dependent  Hall (a) electron and (b) hole mobility in wurtzite GaN as a function of biaxial and uniaxial strain.
  }
\end{figure}

Figure~\ref{figP10} shows the electron and hole Hall mobility of GaN computed for biaxial tensile and uniaxial compressive strains 
of 1\% and of 2\%. 
At room temperature the hole Hall mobility increases from 50~cm$^2$/Vs for the relaxed GaN to 111~cm$^2$/Vs and 119~cm$^2$/Vs for +2\% biaxial and -2\% uniaxial strain, respectively. 
This represents a 230\% increase in hole mobility. 
In contrast, the electron mobility is much less affected by strain.
The results confirm our expectation that, as soon as we change the sign of $\Delta_\text{cf}$, we have an 
enhancement in the hole mobility. 
We emphasize that these results are not sensitive to the details of the calculations
and rests on the \textit{change of ordering} between the split-of band and the light hole and heavy hole bands under applied strain.
We confirmed this finding by performing additional calculations of the variation of $\Delta_\text{cf}$ with strain using the PBE functional, obtaining similar results.

\section{Feasibility of strained GaN}\label{feasibility_gan}

We now investigate the feasibility of realizing high hole mobility GaN experimentally. 
First, we have computed in Fig.~\ref{figP2} the GaN phase diagram and showed that the wurzite structure remains the lowest-enthalpy phase in a large 
pressure and temperature range.
Second, we noticed that biaxial strain of up to 4\% has already been realized experimentally by epitaxial growth on substrates such as AlN or 6H-SiC~\cite{Jain2000,Wagner2002,Li2014d}. 
However, in these experiments the large film thickness induces misfit dislocations~\cite{Floro2004} to release the strain in the sample. 
The dislocations increase defect scattering~\cite{Jena2000}, yielding low hole mobility. 
This may be the reason why high hole mobility GaN has not been observed to date.
Therefore to realize high-hole-mobility GaN we have to devise a plan for preventing dislocation nucleation.

\begin{figure}[t]
  \centering
  \includegraphics[width=0.99\linewidth]{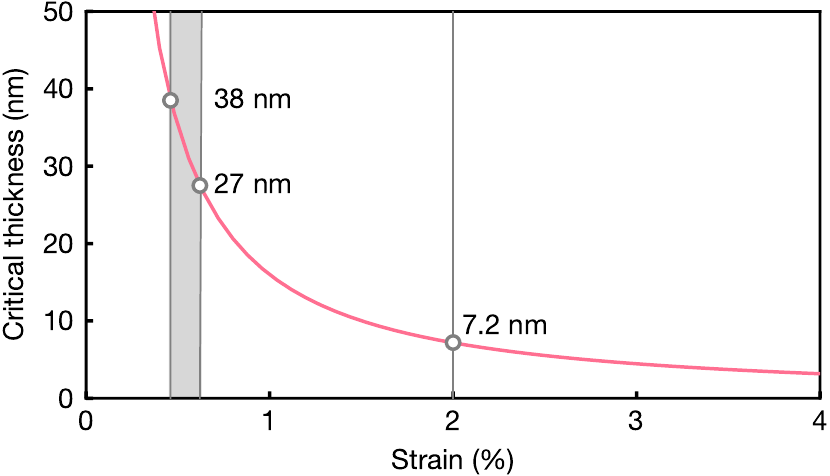}
  \caption{\label{figP11}
   Critical layer thickness of GaN as a function of strain, estimated using Eq.~\eqref{eq:hc}.
   The gray area represents the minimal strain required for crystal-field splitting inversion under uniaxial or biaxial strain.  
  }
\end{figure}

When growing an epilayer on a substrate with a different lattice parameter, dislocations might occur in the epitaxial layer. 
The most common plastic relaxation mechanism is through the formation of misfit dislocations, to accommodate the strain induced by the substrate~\cite{Holec2008}.
The relaxation of misfit strain via plastic flow occurs for an epitaxial layer with a thickness larger than a critical thickness $h_{\rm c}$.
Numerous models have been developed over the years to compute the critical thickness. 
Energy balance models have been developed~\cite{People1985} where the energy of adding a misfit dislocation is balanced with the energy gained by the system from its addition. 
Another popular model developed by Matthews and Blakeslee~\cite{Matthews1974} is based on the force equilibrium method,
in which the forces required to move misfit dislocations are balanced against the elastic stress field due to dislocation interactions. 
Such model was later refined by Fischer~\cite{Fischer1994} using an image-force method where the critical thickness $h_{\rm c}$ for a given strain $\epsilon$ is obtained by solving the following non-linear equation~\cite{Fischer1994} :
\begin{equation}\label{eq:hc}
h_{\rm c} = \frac{b \cos \lambda}{2\epsilon} \bigg[ 1 + \ln \Big(\frac{h_{\rm c}}{b}\Big) \bigg( \frac{1- \nu/4}{4\pi (1+\nu) \cos^2\lambda}  \bigg)  \bigg].
\end{equation}
Here $b=6.026$~bohr is the magnitude of the Burgers vector, $\nu=0.183$ is the experimental Poisson ratio~\cite{Moram2007}, and $\cos \lambda = 0.5$ is the angle between the dislocation Burgers vector and its line direction.
Cracks will typically appears for a film thickness above $h_{\rm c}$~\cite{Cao2007,Dreyer2015}. 
As shown in Fig.~\ref{figP11}, we see that at 2\% strain, films with thickness of up to 7~nm should not exhibit cracks or misfit dislocations.  

It is also possible that the same effect could be achieved using smaller strain levels. 
Indeed, as soon as reversal of the crystal-field splitting is achieved, the hole mobility should significantly increase. 
As discussed in Section~\ref{strain_elBS}, the $sh$ band goes above the $lh$ and $hh$ bands for strain levels above 0.46\% in the case of biaxial tensile strain, 
and above 0.62\% for uniaxial compressive strain.
These values correspond to critical film thicknesses of 38~nm and 27~nm, respectively; as shown in Fig.~\ref{figP11}. 
These values are in agreement with observed critical thicknesses in GaN and AlN, which were found to range between 3 and 30~monolayers depending on the growth temperature~\cite{Sohi2017}.
We also note that such type layer thicknesses have recently become accessible for GaN~\cite{Qi2017,Islam2017}, making our proposal realistic.
In addition, as shown in Table~\ref{table1}, our theoretical approach slightly overestimates the crystal-field splitting with respect to experiment and some theoretical studies. 
As a result, the critical strain required to reverse the crystal-field splitting might be even lower than our estimate.
We emphasize that the engineering of mobility via strain is a common strategy in semiconductors such as Si, Ge, and III-V compounds~\cite{Sun2007,Natarajan2008,Chu2009}, but it has become possible only recently in the case of GaN~\cite{Kim1996,Wagner2002,Rinke2008,Yan2009,Svane2010,Dreyer2013,Horita2017}. 

Finally, an alternative to induce strain via lattice mismatch would be to modify the crystal-field splitting by directly changing the internal parameter $u$, see Fig.~\ref{figP8}(b).
Given that the internal parameter can be tuned by the $A_1$ transverse-optical phonon at $\Gamma$, it should be possible to reverse the crystal-field splitting 
by coherently exciting this optical phonon with femtosecond infrared pulses~\cite{Caviglia2012,Cartella2018}.
This means that we might be able to control the hole mobility in GaN with light instead of strain.

\section{Conclusion}\label{conclu}

In summary, we have computed the phase diagram of GaN and shown that the wurzite phase is the thermodynamical stable phase for a very wide range of pressure and temperatures. 
We have analyzed in detail the electronic bandstructure using many-body corrections and spin-orbit coupling, and showed that the crystal-field splitting heavily depends on the internal parameter of 
the wurtzite structure, and could be tuned via strain engineering. 
%
%
We predicted the room temperature electron and hole Hall mobilities in unstrained GaN to be 1034~cm$^2$/Vs and 52~cm$^2$/Vs, respectively. 
%
%
We showed that the hole mobility can be increased by modifying the ordering of the valence band top such that split-off holes rise 
above the light holes and heavy holes.
This can be achieved using either biaxial tensile strain or uniaxial compressive strain.
We analyzed the effect of strain in GaN including the elastic constants, the high-frequency dielectric constants, Born-effective charges, and phonons.     
We predict over 200\% increase in the hole mobility under strain with respect to the unstrained crystal, reaching values of 120~cm$^2$/Vs under 2\% biaxial tensile or uniaxial compressive strain.
In contrast, the electron mobility remains mostly unaffected. 
We propose to realize such band inversion by reversing the the crystal-field splitting via strain engineering or via optical 
phonon pumping. 
To avoid cracks or misfit dislocations, we propose the use of ultra-thin GaN films (7-40~nm) grown for example by molecular-beam epitaxy on substrates of larger lattice constant than GaN. 
We hope that this work will stimulate further experimental research in high-hole-mobility GaN, and will accelerate progress towards GaN-based CMOS technology and nitride-based high power electronics.

\begin{acknowledgments}
We are grateful to E.~R.~Margine for assistance with the calculation of the band velocity,
and M.~Schlipf for useful discussions. This work was supported by the Leverhulme Trust 
(Grant RL-2012-001), the UK Engineering and Physical Sciences Research Council (grant 
No. EP/M020517/1), the Graphene Flagship (Horizon 2020 Grant No. 785219 - GrapheneCore2),
the University of Oxford Advanced Research Computing (ARC) facility 
(http://dx.doi.org/810.5281/zenodo.22558), the ARCHER UK National Supercomputing Service under 
the AMSEC and CTOA projects, PRACE DECI-13 resource Cartesius at SURFsara, the PRACE DECI-14 resource 
Abel at UiO, and the PRACE-15 and PRACE-17 resources MareNostrum at BSC-CNS. DJ acknowledges support in part from the NSF DMREF award \# 1534303 monitored by Dr. J. Schluter, NSF Award \# 1710298 monitored by Dr. T. Paskova, the NSF CCMR MRSEC Award \#1719875,  AFOSR under Grant
FA9550-17-1-0048 monitored by Dr. K. Goretta, and a research grant from Intel.
\end{acknowledgments}


\end{document}